\documentclass[aps, prl, twocolumn,  letterpaper]{revtex4-2}

\usepackage{graphicx}
\usepackage{amssymb}
\usepackage{amsmath}
\usepackage{float}
\usepackage{txfonts}
\usepackage{bm}
\usepackage{color}
\usepackage[colorlinks=true, linkcolor=blue, anchorcolor=black, citecolor=blue, filecolor=black, menucolor=black, runcolor=black, urlcolor=blue]{hyperref}
\usepackage{mathrsfs}

\begin{document}

\footnotetext[2]{The integer topological invariant is sufficient to characterize braids that correspond to unknots, while additional topological invariants are needed for a complete characterization of knotted braids.
}

\preprint{AIP/123-QED}

\title{Non-Abelian Braiding of Topological Edge Bands}

\author{Yang Long$^1$}
\email{These authors contributed equally to this work.}
\author{Zihao Wang$^1$}
\email{These authors contributed equally to this work.}
\author{Chen Zhang$^2$}
\author{Haoran Xue$^3$}
\author{Yuxin Zhao$^{4,5}$}
\email{Corresponding author: yuxinphy@hku.hk}
\author{Baile Zhang$^{1,6}$}
\email{Corresponding author: blzhang@ntu.edu.sg}
\affiliation{%
$^1$Division of Physics and Applied Physics, School of Physical and Mathematical Sciences, Nanyang Technological University, 21 Nanyang Link, Singapore 637371, Singapore \\
$^2$ National Laboratory of Solid State Microstructures and Department of Physics, Nanjing University, Nanjing 210093, China \\ 
$^3$ Department of Physics, The Chinese University of Hong Kong, Shatin, Hong Kong SAR, China \\
$^4$ Department of Physics and HKU-UCAS Joint Institute for Theoretical and Computational Physics at Hong Kong, The University of Hong Kong, Hong Kong, China \\
$^5$ HK Institute of Quantum Science\& Technology, The University of Hong Kong, Hong Kong, China \\ 
$^6$ Centre for Disruptive Photonic Technologies, Nanyang Technological University, Singapore 637371, Singapore
}%

\date{\today}

\begin{abstract}
Braiding is a geometric concept that manifests itself in a variety of scientific contexts from biology to physics, and has been employed to classify bulk band topology in topological materials. 
Topological edge states can also form braiding structures, as demonstrated recently in a type of topological insulators known as M\"obius insulators, whose topological edge states form two braided bands exhibiting a M\"obius twist. 
While the formation of M\"obius twist is inspiring, it belongs to the simple Abelian braid group $\mathbb{B}_2$. 
The most fascinating features about topological braids rely on the non-Abelianness in the higher-order braid group $\mathbb{B}_N$ ($N \geq 3$),  which necessitates multiple edge bands, but so far it has not been discussed. 
Here, based on the gauge enriched symmetry, we develop a scheme to realize non-Abelian braiding of multiple topological edge bands. 
We propose tight-binding models of topological insulators that are able to generate topological edge states forming non-Abelian braiding structures. 
Experimental demonstrations are conducted in two acoustic crystals, which carry three and four braided acoustic edge bands, respectively. 
The observed braiding structure can correspond to the topological winding in the complex eigenvalue space of projective translation operator, akin to the previously established point-gap winding topology in the bulk of the Hatano-Nelson model. 
Our work also constitutes the realization of non-Abelian braiding topology on an actual crystal platform, but not based on the ``virtual" synthetic dimensions. 
\end{abstract}
                    
\maketitle

Braiding is a multifaceted concept that finds applications across diverse scientific domains, from physics to biology~\cite{Atiyah1990, Leach2004, Kedia2013}. 
Geometrically described by the intricate intertwining of strands forming specific patterns, braiding has significantly broadened its scope in physical systems~\cite{Zhang2022, Pisanty2019, Pisanty2019a, Chen2021, Zhang2023b, Qiu2023}, including the classification of topological bulk bands~\cite{Wang2021, Zhang2023, Hu2021}. Topological edge states can also form braiding structures. For example, in the recently demonstrated M\"obius topological insulators~\cite{Shiozaki2015, Zhang2020, Zhao_2020, Xue2022, Li2022, Jiang2023} , the topological edge states can form two braided edge bands exhibiting a Möbius twist. The time-reversal-invariant M\"obius insulator phase is protected by the projective translation symmetry and is characterized by the $\mathbb{Z}_2$  topological invariant~\cite{ Zhao_2020, Xue2022, Li2022}.

While the presence of M\"obius twist enriches the topological classification, it is worth noting that the Möbius twist represents an Abelian braid formed by two bands, thus falling within the Abelian braid group $\mathbb{B}_2$. 
This prompts a fundamental question: will there be topological insulators beyond the M\"obius topological insulators, namely, the topological insulators whose edge states can form non-Abelian braids of multiple edge bands? 
Note that the non-Abelian braiding topology has sparked some recent experimental progresses in either synthetic dimensions~\cite{Tang2022, Zhang2023b} or electric circuit networks~\cite{Lee2020, Wu2022}, but has not been realized in an actual crystal platform, neither has been applied to topological edge bands. 
In this work, we propose topological insulators protected by translation symmetry, mirror symmetry and time-reversal symmetry, which can realize non-Abelian braiding of topological edge bands.

Conventionally a translation operator translating a crystal by $\bm{a}$ is represented as $e^{i\bm{k}\cdot\bm{a}}$, which exerts no constraint on the $k$-space Hamiltonian. Therefore, translation symmetry is usually not considered as protecting symmetry of topological insulators. However, recent research has shown that this common belief no longer holds when the gauge structure of crystals is considered~\cite{Zhao_2020,kp_PRL_2021,Chen2022,Chen2023}. Particularly, for acoustic crystals gauge structures ubiquitously exist and can be flexibly engineered~\cite{ Xue2022, Li2022, Jiang2023, Yang2015, Ma2019, Zhang2023a,Xue2023, Li2023a}, and therefore it is important to consider symmetry constraints from translation symmetry for topological acoustic crystals~\cite{Xue2018, Xue2019, Xue2020,Xue2021,Xue2022a}.

\begin{figure}[tp!]
\centering
\includegraphics[width=\linewidth]{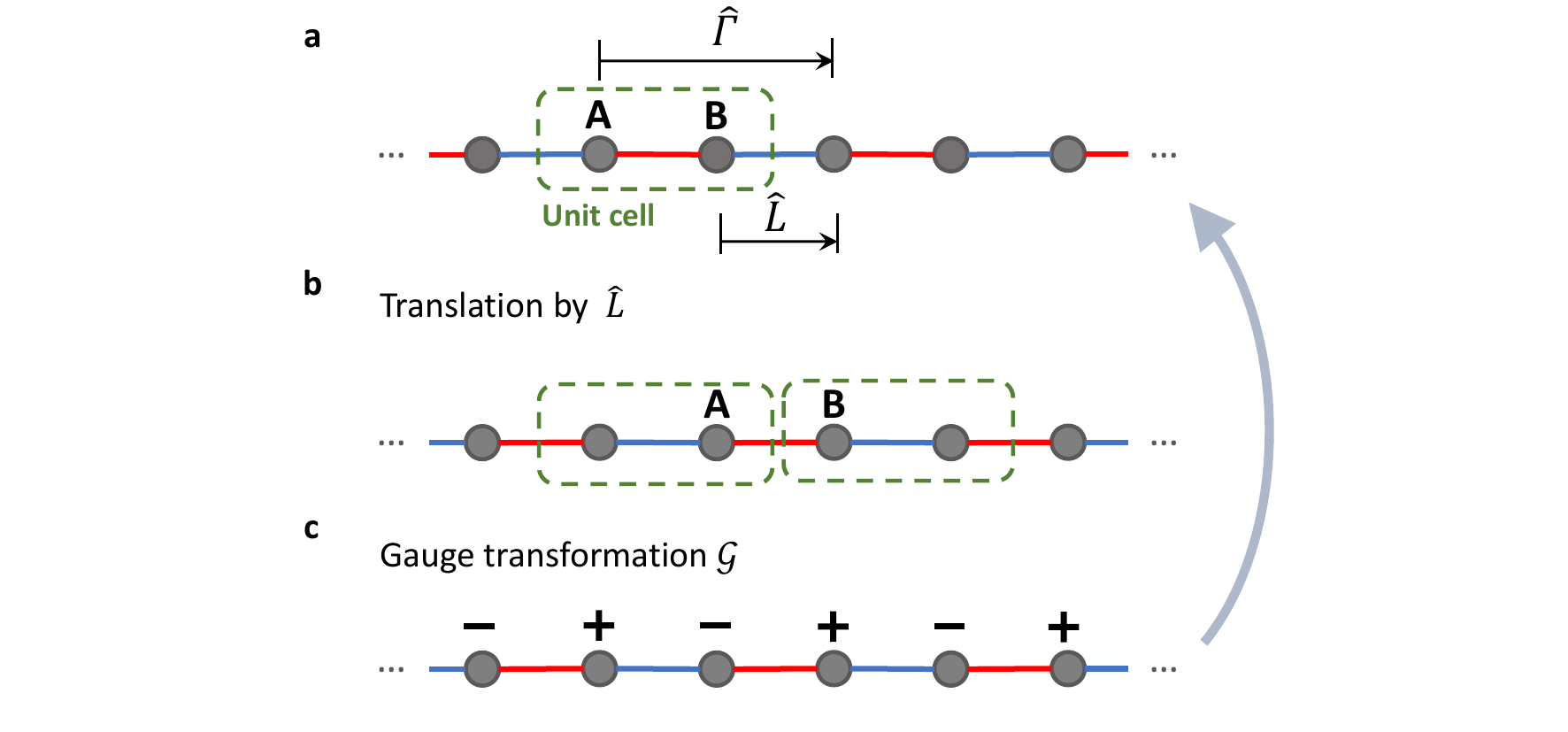}
\caption{\textbf{Illustration for $\mathcal{G}$-dressed translation symmetry}. (a) A 1D chain with a gauge connection configuration $\mathcal{C}$. All hopping amplitudes have real values with the same magnitude. Negative and positive ones are marked in red and blue, respectively. Accordingly, $\mathcal{C}$ is visualized as the blue-red pattern.  $\hat{\Gamma}$ is the unit translation of the primitive unit cells with one surrounded by the dashed line.  (b) The translation of the 1D chain by $\hat{L}$. $\hat{L}$ corresponds to a half of $\hat{\Gamma}$ as indicated in (a). $\mathcal{C}$ in (a) is changed to $\mathcal{C}'$ in (b). (c) The gauge transformation $\mathcal{G}$ on the 1D chain in (b). $\mathcal{G}$ is represented as $\pm$ sign on each site. The gauge transformation switches the color of all bonds as each bond has different signs at its two endpoints. Hence, (b) is transformed to (a) by $\mathcal{G}$.}
\label{fig:projective_symmetry}
\end{figure}

Let us briefly introduce the underlying reason. The gauge connection configuration of a non-interacting tight-binding model consists of the phases of hopping amplitudes [see Fig.~\ref{fig:projective_symmetry}(a) for an example]. Under a given configuration $\mathcal{C}$, we can choose primitive unit cells and the corresponding translation symmetry operators $\hat{\Gamma}_i$ with translation vectors $\bm{a}_i$. Here, the gauge connection configuration $\mathcal{C}$ is invariant under $\hat{\Gamma}_i$. Nevertheless, $\hat{\Gamma}_i$ may not be the minimal translations considering the gauge structure. There may exist a translation operator $\hat{L}_i$ with fractional translation vector $\bm{a}_i/N$ for some $i$, such that $\hat{L}_i$ transforms $\mathcal{C}$ to $\mathcal{C}'$ [see Fig.~\ref{fig:projective_symmetry}(b)], where $\mathcal{C}'$ is related to $\mathcal{C}$ by a gauge transformation $\mathcal{G}_i$ [see Fig.~\ref{fig:projective_symmetry}(c)]. Then, the combination $\hat{\mathcal{L}}_i=\mathcal{G}_i \hat{L}_i$, termed $\mathcal{G}$-dressed translation, is a symmetry of the system [as demonstrated in Fig.~\ref{fig:projective_symmetry}(a), (b) and (c) as a cycle], namely $[H,\hat{\mathcal{L}}_i]=0$. 

We choose the $\mathcal{G}$-dressed translation symmetry $\hat{\mathcal{L}}_x$ as a protecting symmetry of the topological insulators together with mirror and time-reversal symmetries.
In each eigenspace of $\hat{\mathcal{L}}_x$, the topological invariant leads to an edge bands on the edges preserving $\hat{\mathcal{L}}_x$. 
In our work, there are totally $N$ edge bands and each edge band is charged by an eigenvalue of $\hat{\mathcal{L}}_x(k_x)$. 
The $N$ edge bands can be regarded as strands of a braid in the (${\rm Re}[\mathcal{L}_x]$, ${\rm Im}[\mathcal{L}_x]$, $k_x$)-space (${\rm Re/Im}[\cdot]$ denotes the real/imaginary part), which will be denoted as the $\mathcal{L}_x$-$k_x$ space hereafter. 
In the $\mathcal{L}_x$-$k_x$ space, the topological edge states can form a non-Abelian braiding structure. 


\begin{figure}[tp!]
\centering
\includegraphics[width=\linewidth]{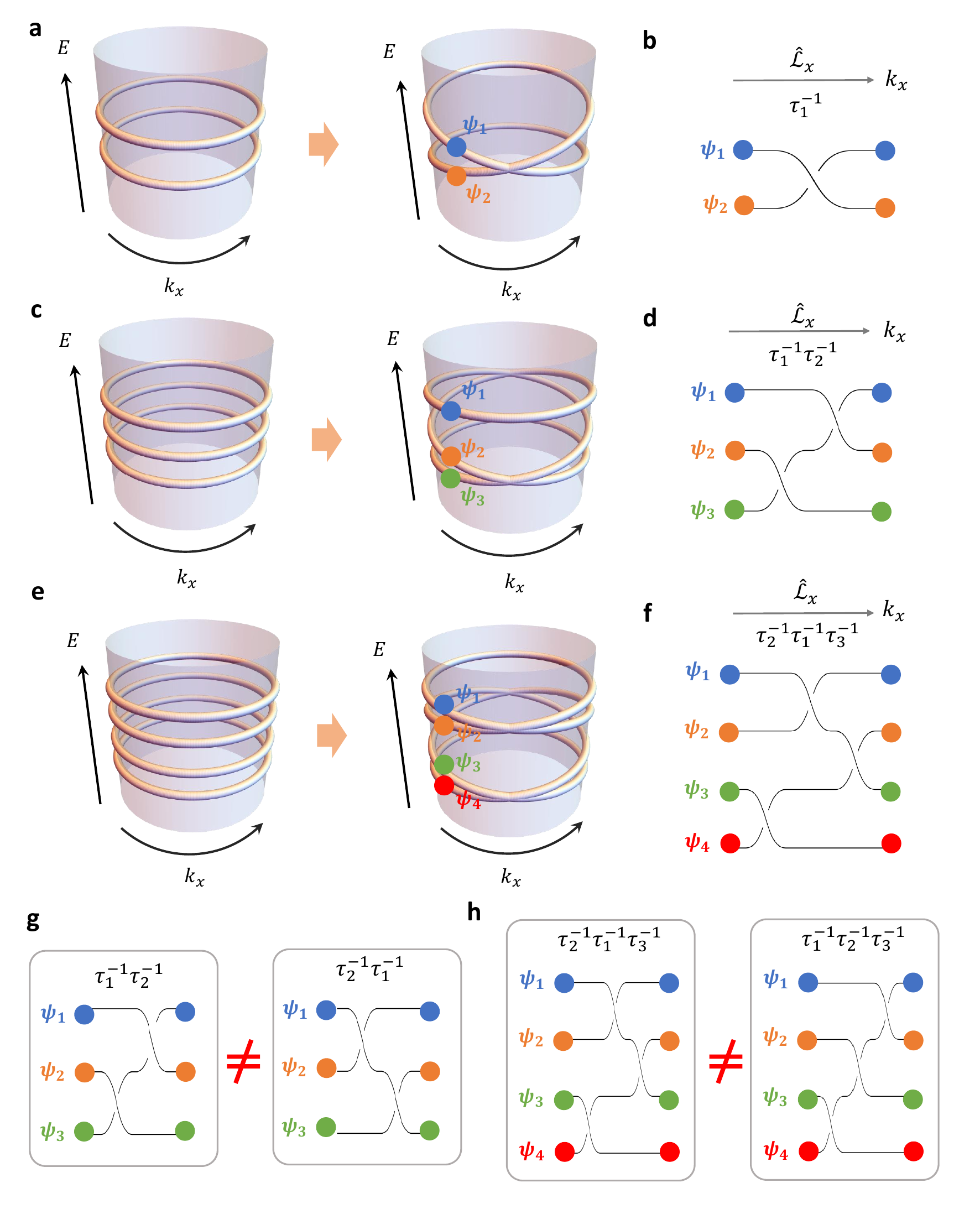}
\caption{{\bf Twisted bandstructures of multiple topological edge bands in the $E$-$k$ space and their braiding in the complex $\mathcal{L}_x$-$k_x$ space. } (a) In the M\"obius topological insulator, two initially separated edge bands can braid with each other after traversing through the edge BZ, forming a M\"obius twist. 
The $\psi_s$ denotes the edge eigenstate. 
Since the momentum $k_x$ is restricted to the edge BZ $k_x \in [-\pi, \pi]$, each topological edge state $\psi_s$ can be represented as a strand of a braid in the $\mathcal{L}_x$-$k_x$ space. (b) The M\"obius twist corresponds to an Abelian braid $\tau_1$ in the $\mathcal{L}_x$-$k_x$ space. 
By exploiting the projective translation symmetry, we can realize the twisted bandstructures of multiple edge bands, such as three edge bands in (c) and four edge bands in (e).  These twisted bandstructures correspond non-Abelian braids in the $\mathcal{L}_x$-$k_x$ space, which are described by $\tau_1^{-1}\tau_2^{-1}$ in (d) and $\tau_2^{-1}\tau_1^{-1}\tau_3^{-1}$ in (f), respectively. (g,h) Non-Abelian properties of $\tau_1^{-1}\tau_2^{-1}$ and $\tau_2^{-1}\tau_1^{-1}\tau_3^{-1}$. }
\label{fig:concept}
\end{figure}

\begin{figure}[tp!]
\centering
\includegraphics[width=\linewidth]{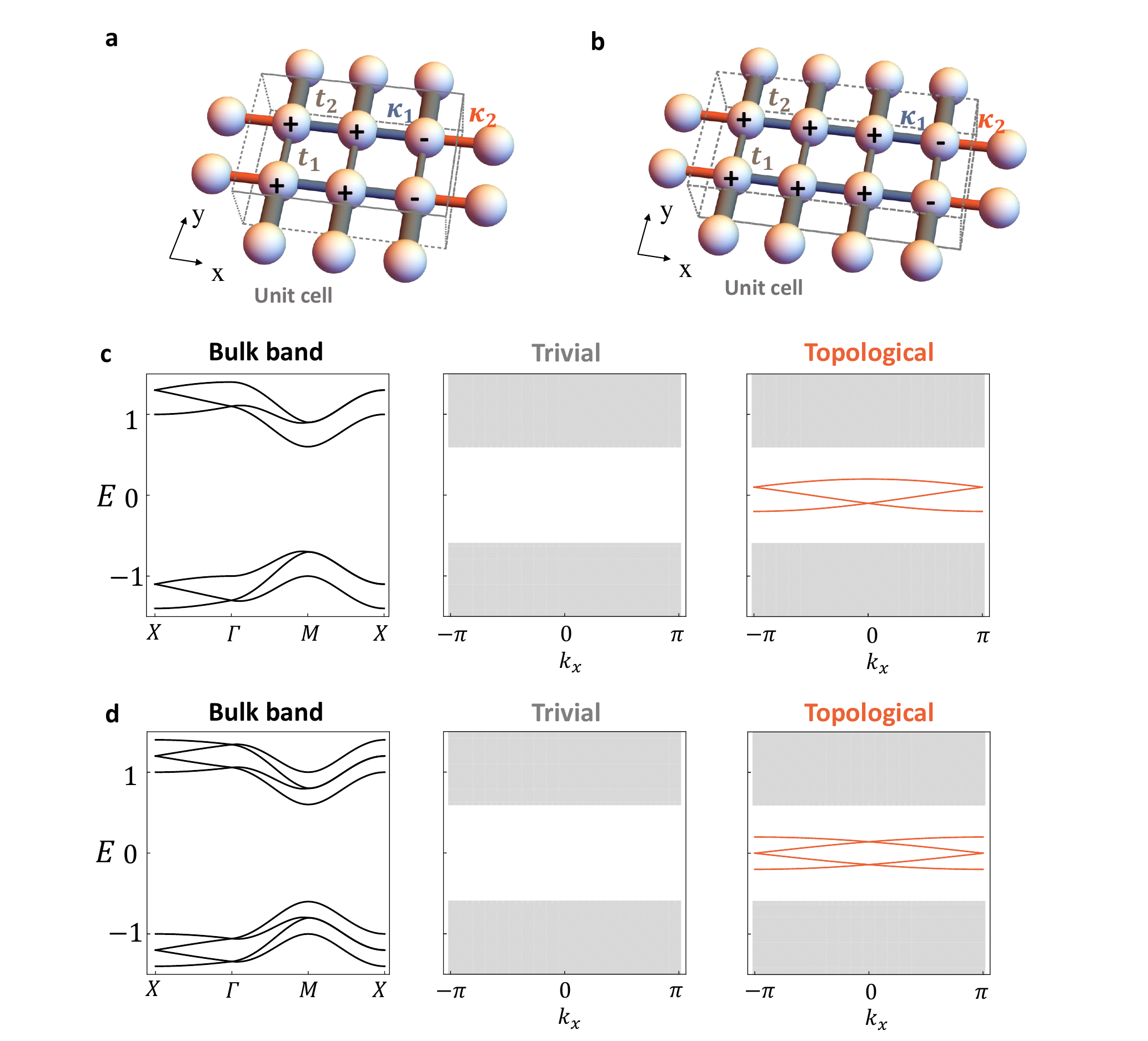}
\caption{{\bf Topological insulators with braided edge bands.} (a) The unit cell of the topological insulator with three braided edge bands. 
Here, ($n,m$) ($n,m \in \mathbb{Z}$) denotes the coordinate of the site. 
The dotted gray box denotes the primitive unit cell. 
(b) The unit cell of the topological insulator with four braided edge bands. Here, $\kappa_1 = -\kappa_2$ due to the projective translation symmetry, where $\kappa_1$, $\kappa_2 \in \mathbb{R}$ and $t_1$, $t_2>0$. (c) The bulk band of the topological insulator in (a) when $t_1 \neq t_2$, $\kappa_1 = -\kappa_2 = -0.1$. The system will open a gap and become topological (trivial) when $t_1=0.2$ and $t_2=1.0$ ($t_1=1.0$ and $t_2=0.2$). 
(d) The bulk band of the topological insulator in (b) when $t_1 \neq t_2$, $\kappa_1 = -\kappa_2 = -0.1$. 
The system will open a gap and become topological (trivial) when $t_1=0.2$ and $t_2=1.0$ ($t_1=1.0$ and $t_2=0.2$). 
The right two figures in (c,d) show the edge spectra, and the gray regions in them denote the bulk bands.}
\label{fig:bandstructure}
\end{figure}

We begin by introducing the twisted energy bandstructures. 
As depicted in Fig.~\ref{fig:concept}(a), recent studies in the M\"obius topological insulator ~\cite{Li2022,Xue2022,Zhao_2020, Li2022} have shown that, under projective translation symmetry, two initially separated edge bands can braid with each other after traversing through the edge BZ, forming a M\"obius twist.  
The topology of the M\"obius twist can be characterized by an Abelian braiding $\tau_1^{-1}$ between two bands in the $\mathcal{L}_x$-$k_x$ space in Fig.~\ref{fig:concept}(b), where $\tau_n$ ($\tau_n^{-1}$) is the braid operation with the meaning that the $n$-th strand crosses under (above) the ($n+1$)-th strand. 
In contrast, our 2D topological insulators can accommodate multiple topological edge states and braid them together, with three bands illustrated in Fig.~\ref{fig:concept}(c) and four bands illustrated in Fig.~\ref{fig:concept}(e), respectively. Now each edge state can return to its original status after traversing through the edge BZ for three or four times. 
These three or four topological edge bands then form non-Abelian braids: $\tau_1^{-1}\tau_2^{-1}$ in Fig.~\ref{fig:concept}(d) and $\tau_2^{-1}\tau_1^{-1}\tau_3^{-1}$ in Fig.~\ref{fig:concept}(f), respectively. 
Namely,  the braids cannot be topologically equivalent after exchanging braid operations, as shown in Fig.~\ref{fig:concept}(g,h). 

To realize the topological edge states depicted in Fig.~\ref{fig:concept}(c,e), we propose the tight-binding models shown in Fig.~\ref{fig:bandstructure}(a,b). 
The Hamiltonians for these models preserve the projective translation symmetry $\hat{\mathcal{L}}_x$, mirror symmetry $\hat{\mathcal{M}}_y$ ($y\rightarrow -y$), and time-reversal symmetry $\hat{\mathcal{T}}$. 
The gauge transformations to be combined with the translation operators $\hat{L}_x$ are given by: $\mathcal{G} = \mathbb{I}_2 \otimes {\rm diag} [1,1,-1]$ in Fig.~\ref{fig:bandstructure}(a) and $\mathcal{G} = \mathbb{I}_2 \otimes {\rm diag} [1,1,1,-1]$ in Fig.~\ref{fig:bandstructure}(b). 
The gauge transformation $\mathcal{G}$ is denoted as the sign $\pm$ on the sites. 
The introductions of the gauge transformations lead to $3\times2$ sites and  $4\times2$ sites for the primitive unit cells, respectively. 
As a result of projective translation symmetry, we have $\kappa_2 = -\kappa_1$ in both cases. 
For the lattice in Fig.~\ref{fig:bandstructure}(a), we can introduce a unitary transformation $U$ to diagonalize $\hat{\mathcal{L}}_x$: $U \hat{\mathcal{L}}_x U^\dagger = e^{i \frac{k_x}{3}} {\rm diag}[-1, e^{i\pi/3}, e^{-i \pi/3}] \otimes \mathbb{I}_2$. 
Since $[\hat{\mathcal{L}}_x, H] = 0$, the Hamiltonian $H$ can be represented in a block diagonalized form $\mathcal{H} = U H U^\dagger = {\rm diag}[h_1, h_2, h_3]$. 
Because $\hat{\mathcal{L}}_x^3 = -e^{i k_x}$, $U$ is not periodic when $k_x$ goes from $-\pi$ to $\pi$ (i.e. $U|_{k_x=-\pi}\neq U|_{k_x=\pi}$), and $\mathcal{H}$ is also not periodic along $k_x$ (i.e. $\mathcal{H}|_{k_x=-\pi}\neq \mathcal{H}|_{k_x=\pi}$). 
We find that the blocks in $\mathcal{H}$ have the following relation~\footnote{
See Supplemental Material at [link] for more details about theoretical analysis about Hamiltonian and topological properties, non-trivial knots or links, experimental discussions, and discussions about the braiding topology in the $\mathcal{L}_x$-$k_x$ space, which include the Ref.~\cite{Jiao2021, Kassel2008}
}:
\begin{equation}
h_{3,2,1} (k_x+2\pi,k_y) = h_{2,1,3} (k_x,k_y).
\label{eq:block_link}
\end{equation}
Here Eq.~\ref{eq:block_link} implies that the blocks can transform into each other in a cyclic manner when $k_x$ goes from $-\pi$ to $\pi$. This gives rise to the 3-fold rotation $\hat{C}_3$ behavior in the basis $\{h_3, h_2, h_1\}$\cite{Note1}. 
Similarly, the lattice in Fig.\ref{fig:bandstructure}(b)  satisfies $h_{4,3,2,1} (k_x+2\pi,k_y) = h_{3,2,1,4} (k_x,k_y)$, leading to the 4-fold rotation $\hat{C}_4$ behavior in the basis of blocks in $\mathcal{H}$~\cite{Note1}. 

\begin{figure}[tp!]
\centering
\includegraphics[width=\linewidth]{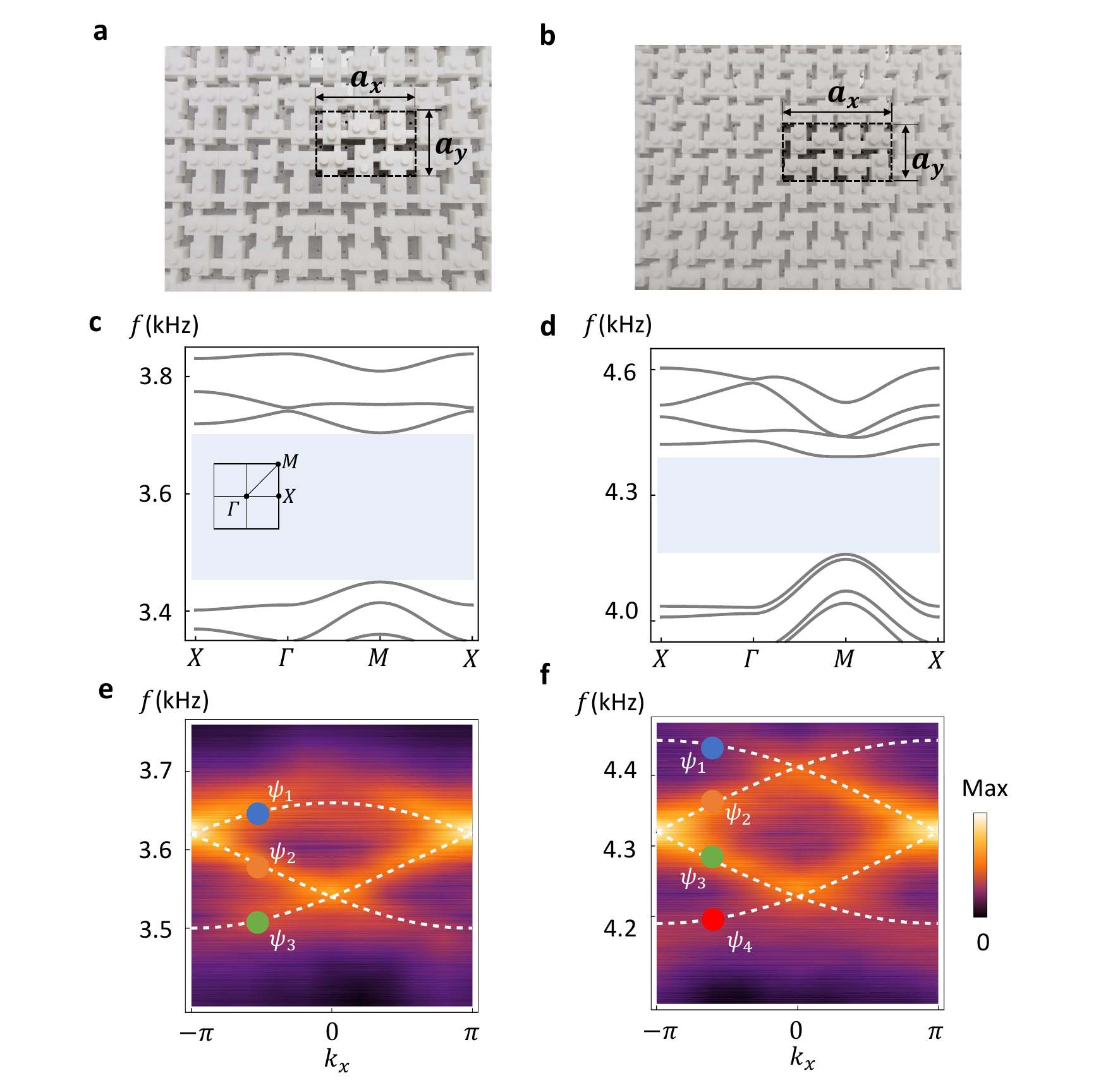}
\caption{{\bf Experimental measurements of twisted edge bandstructures in topological acoustic lattices.} (a) A snapshot of the acoustic crystal corresponding to the tight-binding model in Fig.~\ref{fig:bandstructure}(a). 
The unit cell is denoted in a dashed rectangle. Here, $a_x=180$mm, $a_y=120$mm. 
(b) A snapshot of the acoustic crystal corresponding to the tight-binding model in Fig.~\ref{fig:bandstructure}(b). Here, $a_x=200$mm, $a_y=100$mm. 
(c)  Simulated bulk bandstructure of the lattice in (a). The blue region denote the bulk band gap. 
(d) Simulated bulk bandstructure of the lattice in (b). 
(e) Experimentally measured edge spectrum of the acoustic lattice in (a). 
The background color denotes the Fourier spectrum of the measured pressure field distribution data, and the dotted white lines represent the theoretical predictions with fitted parameters.  
(f) Experimentally measured edge spectrum of the acoustic lattice in (b). 
Here, we use $k_{x,y}$ to denote $k_{x,y}a_{x,y}$, $\psi_n$ denotes the $n$-th edge state. 
}
\label{fig:exp}
\end{figure}

\begin{figure}[tp!]
\centering
\includegraphics[width=\linewidth]{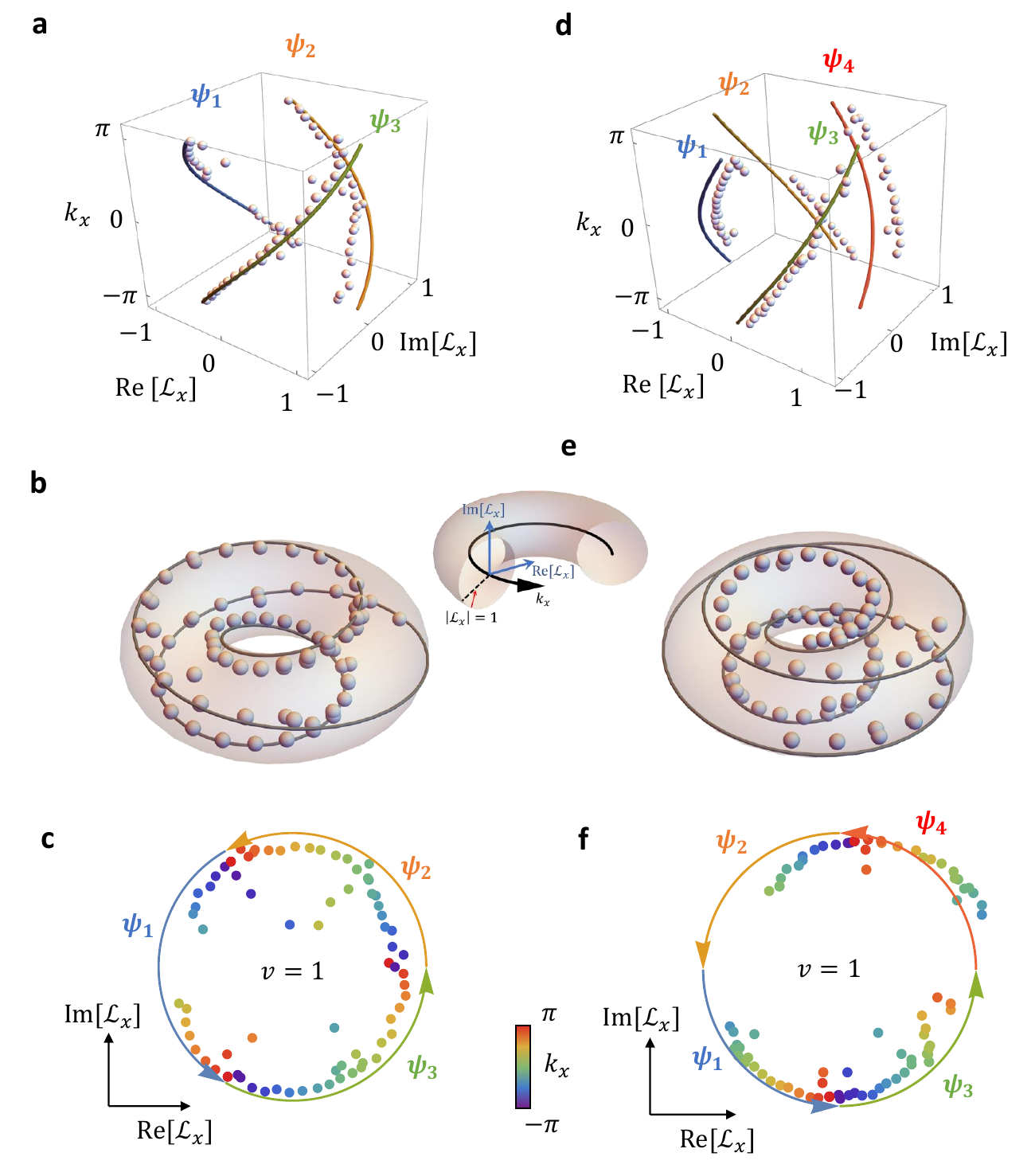}
\caption{{\bf Experimental measurements of non-Abelian braiding of topological edge bands in the $\mathcal{L}_x$-$k_x$ space.} 
(a) Three topological edge bands form a non-Abelian braid $\tau_1^{-1}\tau_2^{-1}$. The lines and spheres denote the theoretical predictions and calculated values based on experimental measurements, respectively. 
The colors of lines correspond to the edge states in Fig.~\ref{fig:exp}(e). 
(b) The knot obtained by presenting the braid in (a) as a torus knot. The gray line represent the theoretical prediction, while the calculated values based on experimental measurements are denoted by the spheres. 
(c) The winding of $\mathcal{L}_x$ in (a) on the (${\rm Re}[\mathcal{L}_x]$, ${\rm Im}[\mathcal{L}_x]$) plane. The positions of points denote the different values of $\mathcal{L}_x$ , with the color representing the wave vector $k_x \in [-\pi, \pi]$.  
The winding behaviour can be represented by a winding number $v=1$. 
(d) Four topological edge bands form a non-Abelian braid $\tau_2^{-1}\tau_1^{-1}\tau_3^{-1}$. 
The colors of lines correspond to the edge states in Fig.~\ref{fig:exp}(f). 
(e) The knot structure obtained by presenting the braid in (d) as a torus knot. 
(f) The winding of $\mathcal{L}_x$ in (d) on the (${\rm Re}[\mathcal{L}_x]$, ${\rm Im}[\mathcal{L}_x]$) plane. The winding behaviour can be represented by a winding number $v=1$. 
 }
\label{fig:braiding}
\end{figure}

With preserved $\hat{\mathcal{L}}_x$, $\hat{\mathcal{M}}_y$ and $\hat{\mathcal{T}}$, the topological properties of lattices in Fig.~\ref{fig:bandstructure}(a,b) can be described by a $\mathbb{Z}$ topological invariant~\cite{Chiu2016, Chiu2013} (i.e., $M\mathbb{Z}$ topology in class AI), which can have topological edge states on an edge protected by $\hat{\mathcal{L}}_x$, similar to the M\"obius insulator~\cite{Zhao_2020,Li2022,Xue2022}. 
Because the preserved $\hat{\mathcal{L}}_x$ can block diagonalize $H$ into $\mathcal{H}$ and the blocks in $\mathcal{H}$ can be transformed from each other by the relation of Eq.~\ref{eq:block_link}, the $\mathbb{Z}$ invariant can be defined according to $h_1$. 
The $M\mathbb{Z}$ topology of $h_1$ can be described by an integer number $N_{-}=N_{-}^{\pi}-N_{-}^0 \in\mathbb{Z}$~\cite{Chiu2013, Note1}, where $N_{-}^{0}$($N_{-}^{\pi}$) denotes the number of occupied bands at $k_y=0$($k_y=\pi$) with $- 1$ eigvenvalues of $\hat{\mathcal{M}}_y$. 
$N_{-}$ can associate with the Zak phase of 1D subsystem $h_1$ when $k_x$ is fixed~\cite{Note1}.  
Note that $N_{-}^{\pi}$ is constant for any $k_x$ because the gap keeps open in the whole BZ. 
When $t_2>t_1$ ($t_1<t_2$), the lattices in Fig.\ref{fig:bandstructure}(a,b) can open a bulk gap in Fig.\ref{fig:bandstructure}(c,d),  respectively, and become topological with $N_{-}=-1$ (trivial with $N_{-}=0$). 
In the topological phase, non-trivial bulk topological invariant ($N_{-}=-1$) of each 1D subsystem $h_i$ can correspond to an in-gap topological edge state on the edge that preserves $\hat{\mathcal{L}}_x$. 
As shown in Fig.\ref{fig:bandstructure}(c,d), multiple in-gap edge states from subsystems $\{h_i\}$ can form the twisted bandstructures, corresponding to interconnected bulk states (e.g., Eq.~\ref{eq:block_link}). 
Because each subsystem $h_i$ belongs to the block with a distinct eigenvalue of $\hat{\mathcal{L}}_x$, these edge states can thus form a non-Abelian braid in the $\mathcal{L}_x$-$k_x$ space (we will discuss in the following).

According to the tight-binding models in Fig.~\ref{fig:bandstructure}(a,b), we design the topological acoustic crystals in Fig.~\ref{fig:exp}(a,b), which exploit coupled acoustic resonators that host dipolar resonances. The positive/negative couplings are implemented by engineering the coupling channels~\cite{Xue2022a}. 
The cuboid resonators used in experiment have dimensions of $48 \text{ mm} \times 24 \text{ mm} \times 12 \text{ mm}$ in Fig.\ref{fig:exp}(a) and $40 \text{ mm} \times 20 \text{ mm} \times 10 \text{ mm}$ in Fig.\ref{fig:exp}(b). 
The resonators are hollow (filled with air) and surrounded by hard walls. The geometrical details can be found in the Supplementary Materials~\cite{Note1}. After obtaining the bandstructures of the acoustic crystals in Fig.\ref{fig:exp}(a,b) numerically, we observe gaps between $3.45\sim3.7$ kHz and $4.16\sim4.4$ kHz, respectively, as shown in Fig.\ref{fig:exp}(c,d). 
We experimentally measure the acoustic pressures of each site on the edge in a finite sample and obtain the edge bands after performing Fourier transformation, as shown in Fig.\ref{fig:exp}(e,f). The experimental results show the presence of topological edge states in the gaps of bulk bands. 
Importantly, these edge states have twisted bandstructures as predicted theoretically in Fig.\ref{fig:bandstructure}(c,d).

In the following, we will discuss the braiding topology of the topological edge bands. 
While the single bulk topological invariant derived from bulk band topology is effective in predicting the presence of topological edge states in two different crsytals in Fig.~\ref{fig:bandstructure}(a,b), the edge bands can form different braids. 
Since $[\hat{\mathcal{L}}_x, H] = 0$, namely $\hat{\mathcal{L}}_x$ shares the same eigenstates as $H$, we can investigate the braiding based on $\hat{\mathcal{L}}_x$. 
To illustrate the braiding structure in Fig.\ref{fig:exp}(e), we numerically simulated the values $\mathcal{L}_x = \langle \psi_s|\hat{\mathcal{L}}_x | \psi_s\rangle$ of each edge state $|\psi_s\rangle$, and also calculated their values based on the measured eigenstates $| \psi_s\rangle$ ~\cite{Note1}, as presented in Fig.\ref{fig:braiding}(a). 
The complex values of $\mathcal{L}_x$ for three edge states in Fig.\ref{fig:exp}(e) form a braiding structure in the $\mathcal{L}_x$-$k_x$ space, which is a non-Abelian braid described by the braid word $\tau_1^{-1}\tau_2^{-1}$ ($\tau_1^{-1}\tau_2^{-1}\neq \tau_2^{-1}\tau_1^{-1}$)~\cite{Note1}. 
Despite these three edge states crossing each other in the $E$-$k$ space, they will satisfy the separable band condition ~\cite{Shen2018} in the $\mathcal{L}_x$-$k_x$ space.  Similarly, in Fig.\ref{fig:braiding}(d), the $\mathcal{L}_x$ values of edge states of the lattice in Fig.~\ref{fig:exp}(b) are simulated numerically and calculated from measured data, forming a non-Abelian braid described by $\tau_2^{-1}\tau_1^{-1}\tau_3^{-1}$~\cite{Note1}.

For three edge bands in Fig.~\ref{fig:braiding}(a), $N=3$, the topology of 3 such separable bands is classified by the braid group $\mathbb{B}_3$. $\mathbb{B}_3$ is isomorphic to the knot group of the trefoil knot $K$: $\pi_1 (\mathbb{R}^3\setminus K)$, which is an infinite non-Abelian group. 
As the momenta at the two ends of the edge BZ are equivalent, the two ends of the braid are identical, thus closing the braid to form a knot or a link. 
Since all bands are connected, the braiding shall form a knot,  as shown in Fig.~\ref{fig:braiding}(b). 
The knot topology in Fig.~\ref{fig:braiding}(b) corresponds to an unknot. 
Because the unknot has the knot group isomorphic to the group of integers $\mathbb{Z}$,  we can exploit an integer topological invariant to describe the topological properties in the $\mathcal{L}_x$-$k_x$ space~\cite{Wang2021, Note2}: 
$v = \frac{1}{2\pi i} \int_0^{2\pi} \partial_{k_x} \ln \det [\mathfrak{L}_x  - \frac{1}{2}{\rm Tr}(\mathfrak{L}_x) ] dk_x$, where $\mathfrak{L}_x$ is a matrix with the element as $(\mathfrak{L}_x)_{m,n} = \langle \psi_m|\hat{\mathcal{L}}_x|\psi_n\rangle$, and $|\psi_n\rangle$ denotes the $n$-th edge state. 
$v$ describes how many times the bands braid as $k_x$ varies from $-\pi$ to $\pi$, with the sign indicating the handedness of the braid, similar to the vorticity of separable bands~\cite{Shen2018}. 
In Fig.~\ref{fig:braiding}(c), we find that $v=1$ for three edge bands in Fig.~\ref{fig:braiding}(a). 
Similarly, for four edge bands ($N=4$), it has an unknot in Fig.~\ref{fig:braiding}(e) and belongs to the $\mathbb{Z}$ group, which has $v=1$, as shown in Fig.~\ref{fig:braiding}(f).

The braiding structures in Fig.~\ref{fig:braiding} are protected by projective translation symmetry $\hat{\mathcal{L}}_x$ and cannot be removed without breaking $\hat{\mathcal{L}}_x$ or closing the band gap~\cite{Note1}. 
The meaning of  $v$ can be understood by mapping to the 1D Hatano-Nelson model within a superlattice~\cite{Note1, Kawabata2019,Gong2018, Bergholtz2021, Hatano1996,Zhang2021}.  
For a large finite system truncated by open boundary conditions, if the edge bands have that $v = 1$, the $\hat{\mathcal{L}}_x$ spectrum of such a finite lattice forms a branch on the (${\rm Re}[\mathcal{L}_x]$, ${\rm Im}[\mathcal{L}_x]$) plane, which is consistent with the previous research works~\cite{Note1, Okuma2020, Wang2021}. 
The non-trivial knots or links of topological edge bands can be realized by introducing long-range couplings~\cite{Note1}. 

\begin{acknowledgments}
In summary, we demonstrate the non-Abelian braiding topology of topological edge states and experimentally observe braiding structures in a topological acoustic crystal. 
Our work opens up a new avenue for exploring braiding topology and can be applied to understand braiding-related topological phenomena~\cite{Zheng2023, Li2023}, such as braids in the isofrequency plane~\cite{Liu2022}, nonequilibrium stochastic dynamics~\cite{Ren2013} or non-Abelian photonics~\cite{Yang2019, Yang2024}.

This research is supported by Singapore National Research Foundation Competitive Research Program Grant No. NRF-CRP23-2019-0007, Singapore Ministry of Education Academic Research Fund Tier 2 Grant No. MOE2019-T2-2-085. 
H.X. acknowledges support from the startup fund of The Chinese University of Hong Kong. 
Y.L. gratefully acknowledges
the support of the Eric and Wendy Schmidt AI in Science
Postdoctoral Fellowship, a Schmidt Futures program. 
\end{acknowledgments}

%

%
%
%
%
%
\bibliography{references}

\begin{thebibliography}{53}%
\makeatletter
\providecommand \@ifxundefined [1]{%
 \@ifx{#1\undefined}
}%
\providecommand \@ifnum [1]{%
 \ifnum #1\expandafter \@firstoftwo
 \else \expandafter \@secondoftwo
 \fi
}%
\providecommand \@ifx [1]{%
 \ifx #1\expandafter \@firstoftwo
 \else \expandafter \@secondoftwo
 \fi
}%
\providecommand \natexlab [1]{#1}%
\providecommand \enquote  [1]{``#1''}%
\providecommand \bibnamefont  [1]{#1}%
\providecommand \bibfnamefont [1]{#1}%
\providecommand \citenamefont [1]{#1}%
\providecommand \href@noop [0]{\@secondoftwo}%
\providecommand \href [0]{\begingroup \@sanitize@url \@href}%
\providecommand \@href[1]{\@@startlink{#1}\@@href}%
\providecommand \@@href[1]{\endgroup#1\@@endlink}%
\providecommand \@sanitize@url [0]{\catcode `\\12\catcode `\$12\catcode
  `\&12\catcode `\#12\catcode `\^12\catcode `\_12\catcode `\%12\relax}%
\providecommand \@@startlink[1]{}%
\providecommand \@@endlink[0]{}%
\providecommand \url  [0]{\begingroup\@sanitize@url \@url }%
\providecommand \@url [1]{\endgroup\@href {#1}{\urlprefix }}%
\providecommand \urlprefix  [0]{URL }%
\providecommand \Eprint [0]{\href }%
\providecommand \doibase [0]{https://doi.org/}%
\providecommand \selectlanguage [0]{\@gobble}%
\providecommand \bibinfo  [0]{\@secondoftwo}%
\providecommand \bibfield  [0]{\@secondoftwo}%
\providecommand \translation [1]{[#1]}%
\providecommand \BibitemOpen [0]{}%
\providecommand \bibitemStop [0]{}%
\providecommand \bibitemNoStop [0]{.\EOS\space}%
\providecommand \EOS [0]{\spacefactor3000\relax}%
\providecommand \BibitemShut  [1]{\csname bibitem#1\endcsname}%
\let\auto@bib@innerbib\@empty
\bibitem [{\citenamefont {Atiyah}(1990)}]{Atiyah1990}%
  \BibitemOpen
  \bibfield  {author} {\bibinfo {author} {\bibfnamefont {M.}~\bibnamefont
  {Atiyah}},\ }\href {https://doi.org/10.1017/cbo9780511623868} {\emph
  {\bibinfo {title} {The Geometry and Physics of Knots}}}\ (\bibinfo
  {publisher} {Cambridge University Press},\ \bibinfo {year}
  {1990})\BibitemShut {NoStop}%
\bibitem [{\citenamefont {Leach}\ \emph {et~al.}(2004)\citenamefont {Leach},
  \citenamefont {Dennis}, \citenamefont {Courtial},\ and\ \citenamefont
  {Padgett}}]{Leach2004}%
  \BibitemOpen
  \bibfield  {author} {\bibinfo {author} {\bibfnamefont {J.}~\bibnamefont
  {Leach}}, \bibinfo {author} {\bibfnamefont {M.~R.}\ \bibnamefont {Dennis}},
  \bibinfo {author} {\bibfnamefont {J.}~\bibnamefont {Courtial}},\ and\
  \bibinfo {author} {\bibfnamefont {M.~J.}\ \bibnamefont {Padgett}},\
  }\bibfield  {title} {\bibinfo {title} {Knotted threads of darkness},\ }\href
  {https://doi.org/10.1038/432165a} {\bibfield  {journal} {\bibinfo  {journal}
  {Nature}\ }\textbf {\bibinfo {volume} {432}},\ \bibinfo {pages} {165}
  (\bibinfo {year} {2004})}\BibitemShut {NoStop}%
\bibitem [{\citenamefont {Kedia}\ \emph {et~al.}(2013)\citenamefont {Kedia},
  \citenamefont {Bialynicki-Birula}, \citenamefont {Peralta-Salas},\ and\
  \citenamefont {Irvine}}]{Kedia2013}%
  \BibitemOpen
  \bibfield  {author} {\bibinfo {author} {\bibfnamefont {H.}~\bibnamefont
  {Kedia}}, \bibinfo {author} {\bibfnamefont {I.}~\bibnamefont
  {Bialynicki-Birula}}, \bibinfo {author} {\bibfnamefont {D.}~\bibnamefont
  {Peralta-Salas}},\ and\ \bibinfo {author} {\bibfnamefont {W.~T.~M.}\
  \bibnamefont {Irvine}},\ }\bibfield  {title} {\bibinfo {title} {Tying knots
  in light fields},\ }\href {https://doi.org/10.1103/physrevlett.111.150404}
  {\bibfield  {journal} {\bibinfo  {journal} {Phys. Rev. Lett.}\ }\textbf
  {\bibinfo {volume} {111}},\ \bibinfo {pages} {150404} (\bibinfo {year}
  {2013})}\BibitemShut {NoStop}%
\bibitem [{\citenamefont {Zhang}\ \emph {et~al.}(2022)\citenamefont {Zhang},
  \citenamefont {Yu}, \citenamefont {Chen}, \citenamefont {Tian}, \citenamefont
  {Chen}, \citenamefont {Sun},\ and\ \citenamefont {Ma}}]{Zhang2022}%
  \BibitemOpen
  \bibfield  {author} {\bibinfo {author} {\bibfnamefont {X.-L.}\ \bibnamefont
  {Zhang}}, \bibinfo {author} {\bibfnamefont {F.}~\bibnamefont {Yu}}, \bibinfo
  {author} {\bibfnamefont {Z.-G.}\ \bibnamefont {Chen}}, \bibinfo {author}
  {\bibfnamefont {Z.-N.}\ \bibnamefont {Tian}}, \bibinfo {author}
  {\bibfnamefont {Q.-D.}\ \bibnamefont {Chen}}, \bibinfo {author}
  {\bibfnamefont {H.-B.}\ \bibnamefont {Sun}},\ and\ \bibinfo {author}
  {\bibfnamefont {G.}~\bibnamefont {Ma}},\ }\bibfield  {title} {\bibinfo
  {title} {Non-abelian braiding on photonic chips},\ }\href
  {https://doi.org/10.1038/s41566-022-00976-2} {\bibfield  {journal} {\bibinfo
  {journal} {Nat. Photonics}\ }\textbf {\bibinfo {volume} {16}},\ \bibinfo
  {pages} {390} (\bibinfo {year} {2022})}\BibitemShut {NoStop}%
\bibitem [{\citenamefont {Pisanty}\ \emph
  {et~al.}(2019{\natexlab{a}})\citenamefont {Pisanty}, \citenamefont {Machado},
  \citenamefont {Vicu{\~{n}}a-Hern{\'{a}}ndez}, \citenamefont {Pic{\'{o}}n},
  \citenamefont {Celi}, \citenamefont {Torres},\ and\ \citenamefont
  {Lewenstein}}]{Pisanty2019}%
  \BibitemOpen
  \bibfield  {author} {\bibinfo {author} {\bibfnamefont {E.}~\bibnamefont
  {Pisanty}}, \bibinfo {author} {\bibfnamefont {G.~J.}\ \bibnamefont
  {Machado}}, \bibinfo {author} {\bibfnamefont {V.}~\bibnamefont
  {Vicu{\~{n}}a-Hern{\'{a}}ndez}}, \bibinfo {author} {\bibfnamefont
  {A.}~\bibnamefont {Pic{\'{o}}n}}, \bibinfo {author} {\bibfnamefont
  {A.}~\bibnamefont {Celi}}, \bibinfo {author} {\bibfnamefont {J.~P.}\
  \bibnamefont {Torres}},\ and\ \bibinfo {author} {\bibfnamefont
  {M.}~\bibnamefont {Lewenstein}},\ }\bibfield  {title} {\bibinfo {title}
  {Knotting fractional-order knots with the polarization state of light},\
  }\href {https://doi.org/10.1038/s41566-019-0450-2} {\bibfield  {journal}
  {\bibinfo  {journal} {Nat. Photonics}\ }\textbf {\bibinfo {volume} {13}},\
  \bibinfo {pages} {569} (\bibinfo {year} {2019}{\natexlab{a}})}\BibitemShut
  {NoStop}%
\bibitem [{\citenamefont {Pisanty}\ \emph
  {et~al.}(2019{\natexlab{b}})\citenamefont {Pisanty}, \citenamefont {Rego},
  \citenamefont {Rom{\'{a}}n}, \citenamefont {Pic{\'{o}}n}, \citenamefont
  {Dorney}, \citenamefont {Kapteyn}, \citenamefont {Murnane}, \citenamefont
  {Plaja}, \citenamefont {Lewenstein},\ and\ \citenamefont
  {Hern{\'{a}}ndez-Garc{\'{\i}}a}}]{Pisanty2019a}%
  \BibitemOpen
  \bibfield  {author} {\bibinfo {author} {\bibfnamefont {E.}~\bibnamefont
  {Pisanty}}, \bibinfo {author} {\bibfnamefont {L.}~\bibnamefont {Rego}},
  \bibinfo {author} {\bibfnamefont {J.~S.}\ \bibnamefont {Rom{\'{a}}n}},
  \bibinfo {author} {\bibfnamefont {A.}~\bibnamefont {Pic{\'{o}}n}}, \bibinfo
  {author} {\bibfnamefont {K.~M.}\ \bibnamefont {Dorney}}, \bibinfo {author}
  {\bibfnamefont {H.~C.}\ \bibnamefont {Kapteyn}}, \bibinfo {author}
  {\bibfnamefont {M.~M.}\ \bibnamefont {Murnane}}, \bibinfo {author}
  {\bibfnamefont {L.}~\bibnamefont {Plaja}}, \bibinfo {author} {\bibfnamefont
  {M.}~\bibnamefont {Lewenstein}},\ and\ \bibinfo {author} {\bibfnamefont
  {C.}~\bibnamefont {Hern{\'{a}}ndez-Garc{\'{\i}}a}},\ }\bibfield  {title}
  {\bibinfo {title} {Conservation of torus-knot angular momentum in high-order
  harmonic generation},\ }\href
  {https://doi.org/10.1103/physrevlett.122.203201} {\bibfield  {journal}
  {\bibinfo  {journal} {Phys. Rev. Lett.}\ }\textbf {\bibinfo {volume} {122}},\
  \bibinfo {pages} {203201} (\bibinfo {year} {2019}{\natexlab{b}})}\BibitemShut
  {NoStop}%
\bibitem [{\citenamefont {Chen}\ \emph {et~al.}(2021)\citenamefont {Chen},
  \citenamefont {Zhang}, \citenamefont {Chan},\ and\ \citenamefont
  {Ma}}]{Chen2021}%
  \BibitemOpen
  \bibfield  {author} {\bibinfo {author} {\bibfnamefont {Z.-G.}\ \bibnamefont
  {Chen}}, \bibinfo {author} {\bibfnamefont {R.-Y.}\ \bibnamefont {Zhang}},
  \bibinfo {author} {\bibfnamefont {C.~T.}\ \bibnamefont {Chan}},\ and\
  \bibinfo {author} {\bibfnamefont {G.}~\bibnamefont {Ma}},\ }\bibfield
  {title} {\bibinfo {title} {Classical non-abelian braiding of acoustic
  modes},\ }\href {https://doi.org/10.1038/s41567-021-01431-9} {\bibfield
  {journal} {\bibinfo  {journal} {Nat. Phys.}\ }\textbf {\bibinfo {volume}
  {18}},\ \bibinfo {pages} {179} (\bibinfo {year} {2021})}\BibitemShut
  {NoStop}%
\bibitem [{\citenamefont {Zhang}\ \emph
  {et~al.}(2023{\natexlab{a}})\citenamefont {Zhang}, \citenamefont {Zhao},
  \citenamefont {Liu}, \citenamefont {Feng}, \citenamefont {Xiong},
  \citenamefont {Wu},\ and\ \citenamefont {Qiu}}]{Zhang2023b}%
  \BibitemOpen
  \bibfield  {author} {\bibinfo {author} {\bibfnamefont {Q.}~\bibnamefont
  {Zhang}}, \bibinfo {author} {\bibfnamefont {L.}~\bibnamefont {Zhao}},
  \bibinfo {author} {\bibfnamefont {X.}~\bibnamefont {Liu}}, \bibinfo {author}
  {\bibfnamefont {X.}~\bibnamefont {Feng}}, \bibinfo {author} {\bibfnamefont
  {L.}~\bibnamefont {Xiong}}, \bibinfo {author} {\bibfnamefont
  {W.}~\bibnamefont {Wu}},\ and\ \bibinfo {author} {\bibfnamefont
  {C.}~\bibnamefont {Qiu}},\ }\bibfield  {title} {\bibinfo {title}
  {Experimental characterization of three-band braid relations in non-hermitian
  acoustic lattices},\ }\href
  {https://doi.org/10.1103/physrevresearch.5.l022050} {\bibfield  {journal}
  {\bibinfo  {journal} {Phys. Rev. Res.}\ }\textbf {\bibinfo {volume} {5}},\
  \bibinfo {pages} {l022050} (\bibinfo {year}
  {2023}{\natexlab{a}})}\BibitemShut {NoStop}%
\bibitem [{\citenamefont {Qiu}\ \emph {et~al.}(2023)\citenamefont {Qiu},
  \citenamefont {Zhang}, \citenamefont {Liu}, \citenamefont {Fan},
  \citenamefont {Zhang},\ and\ \citenamefont {Qiu}}]{Qiu2023}%
  \BibitemOpen
  \bibfield  {author} {\bibinfo {author} {\bibfnamefont {H.}~\bibnamefont
  {Qiu}}, \bibinfo {author} {\bibfnamefont {Q.}~\bibnamefont {Zhang}}, \bibinfo
  {author} {\bibfnamefont {T.}~\bibnamefont {Liu}}, \bibinfo {author}
  {\bibfnamefont {X.}~\bibnamefont {Fan}}, \bibinfo {author} {\bibfnamefont
  {F.}~\bibnamefont {Zhang}},\ and\ \bibinfo {author} {\bibfnamefont
  {C.}~\bibnamefont {Qiu}},\ }\bibfield  {title} {\bibinfo {title} {Minimal
  non-abelian nodal braiding in ideal metamaterials},\ }\href
  {https://doi.org/10.1038/s41467-023-36952-9} {\bibfield  {journal} {\bibinfo
  {journal} {Nat. Commun.}\ }\textbf {\bibinfo {volume} {14}},\ \bibinfo
  {pages} {1261} (\bibinfo {year} {2023})}\BibitemShut {NoStop}%
\bibitem [{\citenamefont {Wang}\ \emph {et~al.}(2021)\citenamefont {Wang},
  \citenamefont {Dutt}, \citenamefont {Wojcik},\ and\ \citenamefont
  {Fan}}]{Wang2021}%
  \BibitemOpen
  \bibfield  {author} {\bibinfo {author} {\bibfnamefont {K.}~\bibnamefont
  {Wang}}, \bibinfo {author} {\bibfnamefont {A.}~\bibnamefont {Dutt}}, \bibinfo
  {author} {\bibfnamefont {C.~C.}\ \bibnamefont {Wojcik}},\ and\ \bibinfo
  {author} {\bibfnamefont {S.}~\bibnamefont {Fan}},\ }\bibfield  {title}
  {\bibinfo {title} {Topological complex-energy braiding of non-hermitian
  bands},\ }\href {https://doi.org/10.1038/s41586-021-03848-x} {\bibfield
  {journal} {\bibinfo  {journal} {Nature}\ }\textbf {\bibinfo {volume} {598}},\
  \bibinfo {pages} {59} (\bibinfo {year} {2021})}\BibitemShut {NoStop}%
\bibitem [{\citenamefont {Zhang}\ \emph
  {et~al.}(2023{\natexlab{b}})\citenamefont {Zhang}, \citenamefont {Li},
  \citenamefont {Sun}, \citenamefont {Liu}, \citenamefont {Zhao}, \citenamefont
  {Feng}, \citenamefont {Fan},\ and\ \citenamefont {Qiu}}]{Zhang2023}%
  \BibitemOpen
  \bibfield  {author} {\bibinfo {author} {\bibfnamefont {Q.}~\bibnamefont
  {Zhang}}, \bibinfo {author} {\bibfnamefont {Y.}~\bibnamefont {Li}}, \bibinfo
  {author} {\bibfnamefont {H.}~\bibnamefont {Sun}}, \bibinfo {author}
  {\bibfnamefont {X.}~\bibnamefont {Liu}}, \bibinfo {author} {\bibfnamefont
  {L.}~\bibnamefont {Zhao}}, \bibinfo {author} {\bibfnamefont {X.}~\bibnamefont
  {Feng}}, \bibinfo {author} {\bibfnamefont {X.}~\bibnamefont {Fan}},\ and\
  \bibinfo {author} {\bibfnamefont {C.}~\bibnamefont {Qiu}},\ }\bibfield
  {title} {\bibinfo {title} {Observation of acoustic non-hermitian bloch braids
  and associated topological phase transitions},\ }\href
  {https://doi.org/10.1103/physrevlett.130.017201} {\bibfield  {journal}
  {\bibinfo  {journal} {Phys. Rev. Lett.}\ }\textbf {\bibinfo {volume} {130}},\
  \bibinfo {pages} {017201} (\bibinfo {year} {2023}{\natexlab{b}})}\BibitemShut
  {NoStop}%
\bibitem [{\citenamefont {Hu}\ and\ \citenamefont {Zhao}(2021)}]{Hu2021}%
  \BibitemOpen
  \bibfield  {author} {\bibinfo {author} {\bibfnamefont {H.}~\bibnamefont
  {Hu}}\ and\ \bibinfo {author} {\bibfnamefont {E.}~\bibnamefont {Zhao}},\
  }\bibfield  {title} {\bibinfo {title} {Knots and non-hermitian bloch bands},\
  }\href {https://doi.org/10.1103/physrevlett.126.010401} {\bibfield  {journal}
  {\bibinfo  {journal} {Phys. Rev. Lett.}\ }\textbf {\bibinfo {volume} {126}},\
  \bibinfo {pages} {010401} (\bibinfo {year} {2021})}\BibitemShut {NoStop}%
\bibitem [{\citenamefont {Shiozaki}\ \emph {et~al.}(2015)\citenamefont
  {Shiozaki}, \citenamefont {Sato},\ and\ \citenamefont {Gomi}}]{Shiozaki2015}%
  \BibitemOpen
  \bibfield  {author} {\bibinfo {author} {\bibfnamefont {K.}~\bibnamefont
  {Shiozaki}}, \bibinfo {author} {\bibfnamefont {M.}~\bibnamefont {Sato}},\
  and\ \bibinfo {author} {\bibfnamefont {K.}~\bibnamefont {Gomi}},\ }\bibfield
  {title} {\bibinfo {title} {$\mathbb{Z}_2$ topology in nonsymmorphic
  crystalline insulators: Möbius twist in surface states},\ }\href
  {https://doi.org/10.1103/physrevb.91.155120} {\bibfield  {journal} {\bibinfo
  {journal} {Phys. Rev. B}\ }\textbf {\bibinfo {volume} {91}},\ \bibinfo
  {pages} {155120} (\bibinfo {year} {2015})}\BibitemShut {NoStop}%
\bibitem [{\citenamefont {Zhang}\ \emph {et~al.}(2020)\citenamefont {Zhang},
  \citenamefont {Wu},\ and\ \citenamefont {Sarma}}]{Zhang2020}%
  \BibitemOpen
  \bibfield  {author} {\bibinfo {author} {\bibfnamefont {R.-X.}\ \bibnamefont
  {Zhang}}, \bibinfo {author} {\bibfnamefont {F.}~\bibnamefont {Wu}},\ and\
  \bibinfo {author} {\bibfnamefont {S.~D.}\ \bibnamefont {Sarma}},\ }\bibfield
  {title} {\bibinfo {title} {Möbius insulator and higher-order topology in
  mnbi2nte3n+1},\ }\href {https://doi.org/10.1103/physrevlett.124.136407}
  {\bibfield  {journal} {\bibinfo  {journal} {Phys. Rev. Lett.}\ }\textbf
  {\bibinfo {volume} {124}},\ \bibinfo {pages} {136407} (\bibinfo {year}
  {2020})}\BibitemShut {NoStop}%
\bibitem [{\citenamefont {Zhao}\ \emph {et~al.}(2020)\citenamefont {Zhao},
  \citenamefont {Huang},\ and\ \citenamefont {Yang}}]{Zhao_2020}%
  \BibitemOpen
  \bibfield  {author} {\bibinfo {author} {\bibfnamefont {Y.~X.}\ \bibnamefont
  {Zhao}}, \bibinfo {author} {\bibfnamefont {Y.-X.}\ \bibnamefont {Huang}},\
  and\ \bibinfo {author} {\bibfnamefont {S.~A.}\ \bibnamefont {Yang}},\
  }\bibfield  {title} {\bibinfo {title} {$\mathbb{Z}_2$-projective
  translational symmetry protected topological phases},\ }\href
  {https://doi.org/10.1103/physrevb.102.161117} {\bibfield  {journal} {\bibinfo
   {journal} {Phy. Rev. B}\ }\textbf {\bibinfo {volume} {102}},\ \bibinfo
  {pages} {161117} (\bibinfo {year} {2020})}\BibitemShut {NoStop}%
\bibitem [{\citenamefont {Xue}\ \emph {et~al.}(2022{\natexlab{a}})\citenamefont
  {Xue}, \citenamefont {Wang}, \citenamefont {Huang}, \citenamefont {Cheng},
  \citenamefont {Yu}, \citenamefont {Foo}, \citenamefont {Zhao}, \citenamefont
  {Yang},\ and\ \citenamefont {Zhang}}]{Xue2022}%
  \BibitemOpen
  \bibfield  {author} {\bibinfo {author} {\bibfnamefont {H.}~\bibnamefont
  {Xue}}, \bibinfo {author} {\bibfnamefont {Z.}~\bibnamefont {Wang}}, \bibinfo
  {author} {\bibfnamefont {Y.-X.}\ \bibnamefont {Huang}}, \bibinfo {author}
  {\bibfnamefont {Z.}~\bibnamefont {Cheng}}, \bibinfo {author} {\bibfnamefont
  {L.}~\bibnamefont {Yu}}, \bibinfo {author} {\bibfnamefont {Y.}~\bibnamefont
  {Foo}}, \bibinfo {author} {\bibfnamefont {Y.}~\bibnamefont {Zhao}}, \bibinfo
  {author} {\bibfnamefont {S.~A.}\ \bibnamefont {Yang}},\ and\ \bibinfo
  {author} {\bibfnamefont {B.}~\bibnamefont {Zhang}},\ }\bibfield  {title}
  {\bibinfo {title} {Projectively enriched symmetry and topology in acoustic
  crystals},\ }\href {https://doi.org/10.1103/physrevlett.128.116802}
  {\bibfield  {journal} {\bibinfo  {journal} {Phys. Rev. Lett.}\ }\textbf
  {\bibinfo {volume} {128}},\ \bibinfo {pages} {116802} (\bibinfo {year}
  {2022}{\natexlab{a}})}\BibitemShut {NoStop}%
\bibitem [{\citenamefont {Li}\ \emph {et~al.}(2022)\citenamefont {Li},
  \citenamefont {Du}, \citenamefont {Zhang}, \citenamefont {Li}, \citenamefont
  {Fan}, \citenamefont {Zhang},\ and\ \citenamefont {Qiu}}]{Li2022}%
  \BibitemOpen
  \bibfield  {author} {\bibinfo {author} {\bibfnamefont {T.}~\bibnamefont
  {Li}}, \bibinfo {author} {\bibfnamefont {J.}~\bibnamefont {Du}}, \bibinfo
  {author} {\bibfnamefont {Q.}~\bibnamefont {Zhang}}, \bibinfo {author}
  {\bibfnamefont {Y.}~\bibnamefont {Li}}, \bibinfo {author} {\bibfnamefont
  {X.}~\bibnamefont {Fan}}, \bibinfo {author} {\bibfnamefont {F.}~\bibnamefont
  {Zhang}},\ and\ \bibinfo {author} {\bibfnamefont {C.}~\bibnamefont {Qiu}},\
  }\bibfield  {title} {\bibinfo {title} {Acoustic möbius insulators from
  projective symmetry},\ }\href
  {https://doi.org/10.1103/physrevlett.128.116803} {\bibfield  {journal}
  {\bibinfo  {journal} {Phys. Rev. Lett.}\ }\textbf {\bibinfo {volume} {128}},\
  \bibinfo {pages} {116803} (\bibinfo {year} {2022})}\BibitemShut {NoStop}%
\bibitem [{\citenamefont {Jiang}\ \emph {et~al.}(2023)\citenamefont {Jiang},
  \citenamefont {Song}, \citenamefont {Li}, \citenamefont {Lu},\ and\
  \citenamefont {Ke}}]{Jiang2023}%
  \BibitemOpen
  \bibfield  {author} {\bibinfo {author} {\bibfnamefont {C.}~\bibnamefont
  {Jiang}}, \bibinfo {author} {\bibfnamefont {Y.}~\bibnamefont {Song}},
  \bibinfo {author} {\bibfnamefont {X.}~\bibnamefont {Li}}, \bibinfo {author}
  {\bibfnamefont {P.}~\bibnamefont {Lu}},\ and\ \bibinfo {author}
  {\bibfnamefont {S.}~\bibnamefont {Ke}},\ }\bibfield  {title} {\bibinfo
  {title} {Photonic möbius topological insulator from projective symmetry in
  multiorbital waveguides},\ }\href {https://doi.org/10.1364/ol.488210}
  {\bibfield  {journal} {\bibinfo  {journal} {Opt. Lett.}\ }\textbf {\bibinfo
  {volume} {48}},\ \bibinfo {pages} {2337} (\bibinfo {year}
  {2023})}\BibitemShut {NoStop}%
\bibitem [{\citenamefont {Tang}\ \emph {et~al.}(2022)\citenamefont {Tang},
  \citenamefont {Ding},\ and\ \citenamefont {Ma}}]{Tang2022}%
  \BibitemOpen
  \bibfield  {author} {\bibinfo {author} {\bibfnamefont {W.}~\bibnamefont
  {Tang}}, \bibinfo {author} {\bibfnamefont {K.}~\bibnamefont {Ding}},\ and\
  \bibinfo {author} {\bibfnamefont {G.}~\bibnamefont {Ma}},\ }\bibfield
  {title} {\bibinfo {title} {Experimental realization of non-abelian
  permutations in a three-state non-hermitian system},\ }\href
  {https://doi.org/10.1093/nsr/nwac010} {\bibfield  {journal} {\bibinfo
  {journal} {Natl. Sci. Rev.}\ }\textbf {\bibinfo {volume} {9}},\ \bibinfo
  {pages} {nwac010} (\bibinfo {year} {2022})}\BibitemShut {NoStop}%
\bibitem [{\citenamefont {Lee}\ \emph {et~al.}(2020)\citenamefont {Lee},
  \citenamefont {Sutrisno}, \citenamefont {Hofmann}, \citenamefont {Helbig},
  \citenamefont {Liu}, \citenamefont {Ang}, \citenamefont {Ang}, \citenamefont
  {Zhang}, \citenamefont {Greiter},\ and\ \citenamefont {Thomale}}]{Lee2020}%
  \BibitemOpen
  \bibfield  {author} {\bibinfo {author} {\bibfnamefont {C.~H.}\ \bibnamefont
  {Lee}}, \bibinfo {author} {\bibfnamefont {A.}~\bibnamefont {Sutrisno}},
  \bibinfo {author} {\bibfnamefont {T.}~\bibnamefont {Hofmann}}, \bibinfo
  {author} {\bibfnamefont {T.}~\bibnamefont {Helbig}}, \bibinfo {author}
  {\bibfnamefont {Y.}~\bibnamefont {Liu}}, \bibinfo {author} {\bibfnamefont
  {Y.~S.}\ \bibnamefont {Ang}}, \bibinfo {author} {\bibfnamefont {L.~K.}\
  \bibnamefont {Ang}}, \bibinfo {author} {\bibfnamefont {X.}~\bibnamefont
  {Zhang}}, \bibinfo {author} {\bibfnamefont {M.}~\bibnamefont {Greiter}},\
  and\ \bibinfo {author} {\bibfnamefont {R.}~\bibnamefont {Thomale}},\
  }\bibfield  {title} {\bibinfo {title} {Imaging nodal knots in momentum space
  through topolectrical circuits},\ }\href
  {https://doi.org/10.1038/s41467-020-17716-1} {\bibfield  {journal} {\bibinfo
  {journal} {Nat. Commun.}\ }\textbf {\bibinfo {volume} {11}},\ \bibinfo
  {pages} {4385} (\bibinfo {year} {2020})}\BibitemShut {NoStop}%
\bibitem [{\citenamefont {Wu}\ \emph {et~al.}(2022)\citenamefont {Wu},
  \citenamefont {Wang}, \citenamefont {Biao}, \citenamefont {Fei},
  \citenamefont {Zhang}, \citenamefont {Yin}, \citenamefont {Hu}, \citenamefont
  {Song}, \citenamefont {Wu}, \citenamefont {Song},\ and\ \citenamefont
  {Yu}}]{Wu2022}%
  \BibitemOpen
  \bibfield  {author} {\bibinfo {author} {\bibfnamefont {J.}~\bibnamefont
  {Wu}}, \bibinfo {author} {\bibfnamefont {Z.}~\bibnamefont {Wang}}, \bibinfo
  {author} {\bibfnamefont {Y.}~\bibnamefont {Biao}}, \bibinfo {author}
  {\bibfnamefont {F.}~\bibnamefont {Fei}}, \bibinfo {author} {\bibfnamefont
  {S.}~\bibnamefont {Zhang}}, \bibinfo {author} {\bibfnamefont
  {Z.}~\bibnamefont {Yin}}, \bibinfo {author} {\bibfnamefont {Y.}~\bibnamefont
  {Hu}}, \bibinfo {author} {\bibfnamefont {Z.}~\bibnamefont {Song}}, \bibinfo
  {author} {\bibfnamefont {T.}~\bibnamefont {Wu}}, \bibinfo {author}
  {\bibfnamefont {F.}~\bibnamefont {Song}},\ and\ \bibinfo {author}
  {\bibfnamefont {R.}~\bibnamefont {Yu}},\ }\bibfield  {title} {\bibinfo
  {title} {Non-abelian gauge fields in circuit systems},\ }\href
  {https://doi.org/10.1038/s41928-022-00833-8} {\bibfield  {journal} {\bibinfo
  {journal} {Nat. Electron.}\ }\textbf {\bibinfo {volume} {5}},\ \bibinfo
  {pages} {635} (\bibinfo {year} {2022})}\BibitemShut {NoStop}%
\bibitem [{\citenamefont {Shao}\ \emph {et~al.}(2021)\citenamefont {Shao},
  \citenamefont {Liu}, \citenamefont {Xiao}, \citenamefont {Yang},\ and\
  \citenamefont {Zhao}}]{kp_PRL_2021}%
  \BibitemOpen
  \bibfield  {author} {\bibinfo {author} {\bibfnamefont {L.~B.}\ \bibnamefont
  {Shao}}, \bibinfo {author} {\bibfnamefont {Q.}~\bibnamefont {Liu}}, \bibinfo
  {author} {\bibfnamefont {R.}~\bibnamefont {Xiao}}, \bibinfo {author}
  {\bibfnamefont {S.~A.}\ \bibnamefont {Yang}},\ and\ \bibinfo {author}
  {\bibfnamefont {Y.~X.}\ \bibnamefont {Zhao}},\ }\bibfield  {title} {\bibinfo
  {title} {Gauge-field extended $k\ifmmode\cdot\else\textperiodcentered\fi{}p$
  method and novel topological phases},\ }\href
  {https://doi.org/10.1103/PhysRevLett.127.076401} {\bibfield  {journal}
  {\bibinfo  {journal} {Phys. Rev. Lett.}\ }\textbf {\bibinfo {volume} {127}},\
  \bibinfo {pages} {076401} (\bibinfo {year} {2021})}\BibitemShut {NoStop}%
\bibitem [{\citenamefont {Chen}\ \emph {et~al.}(2022)\citenamefont {Chen},
  \citenamefont {Yang},\ and\ \citenamefont {Zhao}}]{Chen2022}%
  \BibitemOpen
  \bibfield  {author} {\bibinfo {author} {\bibfnamefont {Z.~Y.}\ \bibnamefont
  {Chen}}, \bibinfo {author} {\bibfnamefont {S.~A.}\ \bibnamefont {Yang}},\
  and\ \bibinfo {author} {\bibfnamefont {Y.~X.}\ \bibnamefont {Zhao}},\
  }\bibfield  {title} {\bibinfo {title} {Brillouin klein bottle from artificial
  gauge fields},\ }\href {https://doi.org/10.1038/s41467-022-29953-7}
  {\bibfield  {journal} {\bibinfo  {journal} {Nat. Commun.}\ }\textbf {\bibinfo
  {volume} {13}},\ \bibinfo {pages} {2215} (\bibinfo {year}
  {2022})}\BibitemShut {NoStop}%
\bibitem [{\citenamefont {Chen}\ \emph {et~al.}(2023)\citenamefont {Chen},
  \citenamefont {Zhang}, \citenamefont {Yang},\ and\ \citenamefont
  {Zhao}}]{Chen2023}%
  \BibitemOpen
  \bibfield  {author} {\bibinfo {author} {\bibfnamefont {Z.~Y.}\ \bibnamefont
  {Chen}}, \bibinfo {author} {\bibfnamefont {Z.}~\bibnamefont {Zhang}},
  \bibinfo {author} {\bibfnamefont {S.~A.}\ \bibnamefont {Yang}},\ and\
  \bibinfo {author} {\bibfnamefont {Y.~X.}\ \bibnamefont {Zhao}},\ }\bibfield
  {title} {\bibinfo {title} {Classification of time-reversal-invariant crystals
  with gauge structures},\ }\href {https://doi.org/10.1038/s41467-023-36447-7}
  {\bibfield  {journal} {\bibinfo  {journal} {Nat. Commun.}\ }\textbf {\bibinfo
  {volume} {14}},\ \bibinfo {pages} {743} (\bibinfo {year} {2023})}\BibitemShut
  {NoStop}%
\bibitem [{\citenamefont {Yang}\ \emph {et~al.}(2015)\citenamefont {Yang},
  \citenamefont {Gao}, \citenamefont {Shi}, \citenamefont {Lin}, \citenamefont
  {Gao}, \citenamefont {Chong},\ and\ \citenamefont {Zhang}}]{Yang2015}%
  \BibitemOpen
  \bibfield  {author} {\bibinfo {author} {\bibfnamefont {Z.}~\bibnamefont
  {Yang}}, \bibinfo {author} {\bibfnamefont {F.}~\bibnamefont {Gao}}, \bibinfo
  {author} {\bibfnamefont {X.}~\bibnamefont {Shi}}, \bibinfo {author}
  {\bibfnamefont {X.}~\bibnamefont {Lin}}, \bibinfo {author} {\bibfnamefont
  {Z.}~\bibnamefont {Gao}}, \bibinfo {author} {\bibfnamefont {Y.}~\bibnamefont
  {Chong}},\ and\ \bibinfo {author} {\bibfnamefont {B.}~\bibnamefont {Zhang}},\
  }\bibfield  {title} {\bibinfo {title} {Topological acoustics},\ }\href
  {https://doi.org/10.1103/physrevlett.114.114301} {\bibfield  {journal}
  {\bibinfo  {journal} {Phys. Rev. Lett.}\ }\textbf {\bibinfo {volume} {114}},\
  \bibinfo {pages} {114301} (\bibinfo {year} {2015})}\BibitemShut {NoStop}%
\bibitem [{\citenamefont {Ma}\ \emph {et~al.}(2019)\citenamefont {Ma},
  \citenamefont {Xiao},\ and\ \citenamefont {Chan}}]{Ma2019}%
  \BibitemOpen
  \bibfield  {author} {\bibinfo {author} {\bibfnamefont {G.}~\bibnamefont
  {Ma}}, \bibinfo {author} {\bibfnamefont {M.}~\bibnamefont {Xiao}},\ and\
  \bibinfo {author} {\bibfnamefont {C.~T.}\ \bibnamefont {Chan}},\ }\bibfield
  {title} {\bibinfo {title} {Topological phases in acoustic and mechanical
  systems},\ }\href {https://doi.org/10.1038/s42254-019-0030-x} {\bibfield
  {journal} {\bibinfo  {journal} {Nat. Rev. Phys.}\ }\textbf {\bibinfo {volume}
  {1}},\ \bibinfo {pages} {281} (\bibinfo {year} {2019})}\BibitemShut {NoStop}%
\bibitem [{\citenamefont {Zhang}\ \emph
  {et~al.}(2023{\natexlab{c}})\citenamefont {Zhang}, \citenamefont
  {Zangeneh-Nejad}, \citenamefont {Chen}, \citenamefont {Lu},\ and\
  \citenamefont {Christensen}}]{Zhang2023a}%
  \BibitemOpen
  \bibfield  {author} {\bibinfo {author} {\bibfnamefont {X.}~\bibnamefont
  {Zhang}}, \bibinfo {author} {\bibfnamefont {F.}~\bibnamefont
  {Zangeneh-Nejad}}, \bibinfo {author} {\bibfnamefont {Z.-G.}\ \bibnamefont
  {Chen}}, \bibinfo {author} {\bibfnamefont {M.-H.}\ \bibnamefont {Lu}},\ and\
  \bibinfo {author} {\bibfnamefont {J.}~\bibnamefont {Christensen}},\
  }\bibfield  {title} {\bibinfo {title} {A second wave of topological phenomena
  in photonics and acoustics},\ }\href
  {https://doi.org/10.1038/s41586-023-06163-9} {\bibfield  {journal} {\bibinfo
  {journal} {Nature}\ }\textbf {\bibinfo {volume} {618}},\ \bibinfo {pages}
  {687} (\bibinfo {year} {2023}{\natexlab{c}})}\BibitemShut {NoStop}%
\bibitem [{\citenamefont {Xue}\ \emph {et~al.}(2023)\citenamefont {Xue},
  \citenamefont {Chen}, \citenamefont {Cheng}, \citenamefont {Dai},
  \citenamefont {Long}, \citenamefont {Zhao},\ and\ \citenamefont
  {Zhang}}]{Xue2023}%
  \BibitemOpen
  \bibfield  {author} {\bibinfo {author} {\bibfnamefont {H.}~\bibnamefont
  {Xue}}, \bibinfo {author} {\bibfnamefont {Z.~Y.}\ \bibnamefont {Chen}},
  \bibinfo {author} {\bibfnamefont {Z.}~\bibnamefont {Cheng}}, \bibinfo
  {author} {\bibfnamefont {J.~X.}\ \bibnamefont {Dai}}, \bibinfo {author}
  {\bibfnamefont {Y.}~\bibnamefont {Long}}, \bibinfo {author} {\bibfnamefont
  {Y.~X.}\ \bibnamefont {Zhao}},\ and\ \bibinfo {author} {\bibfnamefont
  {B.}~\bibnamefont {Zhang}},\ }\bibfield  {title} {\bibinfo {title}
  {Stiefel-whitney topological charges in a three-dimensional acoustic
  nodal-line crystal},\ }\href {https://doi.org/10.1038/s41467-023-40252-7}
  {\bibfield  {journal} {\bibinfo  {journal} {Nat. Commun.}\ }\textbf {\bibinfo
  {volume} {14}},\ \bibinfo {pages} {4563} (\bibinfo {year}
  {2023})}\BibitemShut {NoStop}%
\bibitem [{\citenamefont {Li}\ \emph {et~al.}(2023{\natexlab{a}})\citenamefont
  {Li}, \citenamefont {Liu}, \citenamefont {Zhang},\ and\ \citenamefont
  {Qiu}}]{Li2023a}%
  \BibitemOpen
  \bibfield  {author} {\bibinfo {author} {\bibfnamefont {T.}~\bibnamefont
  {Li}}, \bibinfo {author} {\bibfnamefont {L.}~\bibnamefont {Liu}}, \bibinfo
  {author} {\bibfnamefont {Q.}~\bibnamefont {Zhang}},\ and\ \bibinfo {author}
  {\bibfnamefont {C.}~\bibnamefont {Qiu}},\ }\bibfield  {title} {\bibinfo
  {title} {Acoustic realization of projective mirror chern insulators},\ }\href
  {https://doi.org/10.1038/s42005-023-01393-9} {\bibfield  {journal} {\bibinfo
  {journal} {Commun. Phys.}\ }\textbf {\bibinfo {volume} {6}},\ \bibinfo
  {pages} {268} (\bibinfo {year} {2023}{\natexlab{a}})}\BibitemShut {NoStop}%
\bibitem [{\citenamefont {Xue}\ \emph {et~al.}(2018)\citenamefont {Xue},
  \citenamefont {Yang}, \citenamefont {Gao}, \citenamefont {Chong},\ and\
  \citenamefont {Zhang}}]{Xue2018}%
  \BibitemOpen
  \bibfield  {author} {\bibinfo {author} {\bibfnamefont {H.}~\bibnamefont
  {Xue}}, \bibinfo {author} {\bibfnamefont {Y.}~\bibnamefont {Yang}}, \bibinfo
  {author} {\bibfnamefont {F.}~\bibnamefont {Gao}}, \bibinfo {author}
  {\bibfnamefont {Y.}~\bibnamefont {Chong}},\ and\ \bibinfo {author}
  {\bibfnamefont {B.}~\bibnamefont {Zhang}},\ }\bibfield  {title} {\bibinfo
  {title} {Acoustic higher-order topological insulator on a kagome lattice},\
  }\href {https://doi.org/10.1038/s41563-018-0251-x} {\bibfield  {journal}
  {\bibinfo  {journal} {Nat. Mater.}\ }\textbf {\bibinfo {volume} {18}},\
  \bibinfo {pages} {108} (\bibinfo {year} {2018})}\BibitemShut {NoStop}%
\bibitem [{\citenamefont {Xue}\ \emph {et~al.}(2019)\citenamefont {Xue},
  \citenamefont {Yang}, \citenamefont {Liu}, \citenamefont {Gao}, \citenamefont
  {Chong},\ and\ \citenamefont {Zhang}}]{Xue2019}%
  \BibitemOpen
  \bibfield  {author} {\bibinfo {author} {\bibfnamefont {H.}~\bibnamefont
  {Xue}}, \bibinfo {author} {\bibfnamefont {Y.}~\bibnamefont {Yang}}, \bibinfo
  {author} {\bibfnamefont {G.}~\bibnamefont {Liu}}, \bibinfo {author}
  {\bibfnamefont {F.}~\bibnamefont {Gao}}, \bibinfo {author} {\bibfnamefont
  {Y.}~\bibnamefont {Chong}},\ and\ \bibinfo {author} {\bibfnamefont
  {B.}~\bibnamefont {Zhang}},\ }\bibfield  {title} {\bibinfo {title}
  {Realization of an acoustic third-order topological insulator},\ }\href
  {https://doi.org/10.1103/physrevlett.122.244301} {\bibfield  {journal}
  {\bibinfo  {journal} {Phys. Rev. Lett.}\ }\textbf {\bibinfo {volume} {122}},\
  \bibinfo {pages} {244301} (\bibinfo {year} {2019})}\BibitemShut {NoStop}%
\bibitem [{\citenamefont {Xue}\ \emph {et~al.}(2020)\citenamefont {Xue},
  \citenamefont {Ge}, \citenamefont {Sun}, \citenamefont {Wang}, \citenamefont
  {Jia}, \citenamefont {Guan}, \citenamefont {Yuan}, \citenamefont {Chong},\
  and\ \citenamefont {Zhang}}]{Xue2020}%
  \BibitemOpen
  \bibfield  {author} {\bibinfo {author} {\bibfnamefont {H.}~\bibnamefont
  {Xue}}, \bibinfo {author} {\bibfnamefont {Y.}~\bibnamefont {Ge}}, \bibinfo
  {author} {\bibfnamefont {H.-X.}\ \bibnamefont {Sun}}, \bibinfo {author}
  {\bibfnamefont {Q.}~\bibnamefont {Wang}}, \bibinfo {author} {\bibfnamefont
  {D.}~\bibnamefont {Jia}}, \bibinfo {author} {\bibfnamefont {Y.-J.}\
  \bibnamefont {Guan}}, \bibinfo {author} {\bibfnamefont {S.-Q.}\ \bibnamefont
  {Yuan}}, \bibinfo {author} {\bibfnamefont {Y.}~\bibnamefont {Chong}},\ and\
  \bibinfo {author} {\bibfnamefont {B.}~\bibnamefont {Zhang}},\ }\bibfield
  {title} {\bibinfo {title} {Observation of an acoustic octupole topological
  insulator},\ }\href {https://doi.org/10.1038/s41467-020-16350-1} {\bibfield
  {journal} {\bibinfo  {journal} {Nat. Commun.}\ }\textbf {\bibinfo {volume}
  {11}},\ \bibinfo {pages} {2442} (\bibinfo {year} {2020})}\BibitemShut
  {NoStop}%
\bibitem [{\citenamefont {Xue}\ \emph {et~al.}(2021)\citenamefont {Xue},
  \citenamefont {Jia}, \citenamefont {Ge}, \citenamefont {jun Guan},
  \citenamefont {Wang}, \citenamefont {qi~Yuan}, \citenamefont {xiang Sun},
  \citenamefont {Chong},\ and\ \citenamefont {Zhang}}]{Xue2021}%
  \BibitemOpen
  \bibfield  {author} {\bibinfo {author} {\bibfnamefont {H.}~\bibnamefont
  {Xue}}, \bibinfo {author} {\bibfnamefont {D.}~\bibnamefont {Jia}}, \bibinfo
  {author} {\bibfnamefont {Y.}~\bibnamefont {Ge}}, \bibinfo {author}
  {\bibfnamefont {Y.}~\bibnamefont {jun Guan}}, \bibinfo {author}
  {\bibfnamefont {Q.}~\bibnamefont {Wang}}, \bibinfo {author} {\bibfnamefont
  {S.}~\bibnamefont {qi~Yuan}}, \bibinfo {author} {\bibfnamefont
  {H.}~\bibnamefont {xiang Sun}}, \bibinfo {author} {\bibfnamefont
  {Y.}~\bibnamefont {Chong}},\ and\ \bibinfo {author} {\bibfnamefont
  {B.}~\bibnamefont {Zhang}},\ }\bibfield  {title} {\bibinfo {title}
  {Observation of dislocation-induced topological modes in a three-dimensional
  acoustic topological insulator},\ }\href
  {https://doi.org/10.1103/physrevlett.127.214301} {\bibfield  {journal}
  {\bibinfo  {journal} {Phys. Rev. Lett.}\ }\textbf {\bibinfo {volume} {127}},\
  \bibinfo {pages} {214301} (\bibinfo {year} {2021})}\BibitemShut {NoStop}%
\bibitem [{\citenamefont {Xue}\ \emph {et~al.}(2022{\natexlab{b}})\citenamefont
  {Xue}, \citenamefont {Yang},\ and\ \citenamefont {Zhang}}]{Xue2022a}%
  \BibitemOpen
  \bibfield  {author} {\bibinfo {author} {\bibfnamefont {H.}~\bibnamefont
  {Xue}}, \bibinfo {author} {\bibfnamefont {Y.}~\bibnamefont {Yang}},\ and\
  \bibinfo {author} {\bibfnamefont {B.}~\bibnamefont {Zhang}},\ }\bibfield
  {title} {\bibinfo {title} {Topological acoustics},\ }\href
  {https://doi.org/10.1038/s41578-022-00465-6} {\bibfield  {journal} {\bibinfo
  {journal} {Nat. Rev. Mater.}\ }\textbf {\bibinfo {volume} {7}},\ \bibinfo
  {pages} {974} (\bibinfo {year} {2022}{\natexlab{b}})}\BibitemShut {NoStop}%
\bibitem [{Note1()}]{Note1}%
  \BibitemOpen
  \bibinfo {note} {See Supplemental Material at [link] for more details about
  theoretical analysis about Hamiltonian and topological properties,
  non-trivial knots or links, experimental discussions, and discussions about
  the braiding topology in the $\protect \mathcal {L}_x$-$k_x$ space, which
  include the Ref.~\cite {Jiao2021, Kassel2008}}\BibitemShut {NoStop}%
\bibitem [{\citenamefont {Chiu}\ \emph {et~al.}(2016)\citenamefont {Chiu},
  \citenamefont {Teo}, \citenamefont {Schnyder},\ and\ \citenamefont
  {Ryu}}]{Chiu2016}%
  \BibitemOpen
  \bibfield  {author} {\bibinfo {author} {\bibfnamefont {C.-K.}\ \bibnamefont
  {Chiu}}, \bibinfo {author} {\bibfnamefont {J.~C.}\ \bibnamefont {Teo}},
  \bibinfo {author} {\bibfnamefont {A.~P.}\ \bibnamefont {Schnyder}},\ and\
  \bibinfo {author} {\bibfnamefont {S.}~\bibnamefont {Ryu}},\ }\bibfield
  {title} {\bibinfo {title} {Classification of topological quantum matter with
  symmetries},\ }\href {https://doi.org/10.1103/revmodphys.88.035005}
  {\bibfield  {journal} {\bibinfo  {journal} {Rev. Mod. Phys.}\ }\textbf
  {\bibinfo {volume} {88}},\ \bibinfo {pages} {035005} (\bibinfo {year}
  {2016})}\BibitemShut {NoStop}%
\bibitem [{\citenamefont {Chiu}\ \emph {et~al.}(2013)\citenamefont {Chiu},
  \citenamefont {Yao},\ and\ \citenamefont {Ryu}}]{Chiu2013}%
  \BibitemOpen
  \bibfield  {author} {\bibinfo {author} {\bibfnamefont {C.-K.}\ \bibnamefont
  {Chiu}}, \bibinfo {author} {\bibfnamefont {H.}~\bibnamefont {Yao}},\ and\
  \bibinfo {author} {\bibfnamefont {S.}~\bibnamefont {Ryu}},\ }\bibfield
  {title} {\bibinfo {title} {Classification of topological insulators and
  superconductors in the presence of reflection symmetry},\ }\href
  {https://doi.org/10.1103/physrevb.88.075142} {\bibfield  {journal} {\bibinfo
  {journal} {Phys. Rev. B}\ }\textbf {\bibinfo {volume} {88}},\ \bibinfo
  {pages} {075142} (\bibinfo {year} {2013})}\BibitemShut {NoStop}%
\bibitem [{\citenamefont {Shen}\ \emph {et~al.}(2018)\citenamefont {Shen},
  \citenamefont {Zhen},\ and\ \citenamefont {Fu}}]{Shen2018}%
  \BibitemOpen
  \bibfield  {author} {\bibinfo {author} {\bibfnamefont {H.}~\bibnamefont
  {Shen}}, \bibinfo {author} {\bibfnamefont {B.}~\bibnamefont {Zhen}},\ and\
  \bibinfo {author} {\bibfnamefont {L.}~\bibnamefont {Fu}},\ }\bibfield
  {title} {\bibinfo {title} {Topological band theory for non-hermitian
  hamiltonians},\ }\href {https://doi.org/10.1103/physrevlett.120.146402}
  {\bibfield  {journal} {\bibinfo  {journal} {Phys. Rev. Lett.}\ }\textbf
  {\bibinfo {volume} {120}},\ \bibinfo {pages} {146402} (\bibinfo {year}
  {2018})}\BibitemShut {NoStop}%
\bibitem [{Note2()}]{Note2}%
  \BibitemOpen
  \bibinfo {note} {The integer topological invariant is sufficient to
  characterize braids that correspond to unknots, while additional topological
  invariants are needed for a complete characterization of knotted
  braids.}\BibitemShut {Stop}%
\bibitem [{\citenamefont {Kawabata}\ \emph {et~al.}(2019)\citenamefont
  {Kawabata}, \citenamefont {Shiozaki}, \citenamefont {Ueda},\ and\
  \citenamefont {Sato}}]{Kawabata2019}%
  \BibitemOpen
  \bibfield  {author} {\bibinfo {author} {\bibfnamefont {K.}~\bibnamefont
  {Kawabata}}, \bibinfo {author} {\bibfnamefont {K.}~\bibnamefont {Shiozaki}},
  \bibinfo {author} {\bibfnamefont {M.}~\bibnamefont {Ueda}},\ and\ \bibinfo
  {author} {\bibfnamefont {M.}~\bibnamefont {Sato}},\ }\bibfield  {title}
  {\bibinfo {title} {Symmetry and topology in non-hermitian physics},\ }\href
  {https://doi.org/10.1103/physrevx.9.041015} {\bibfield  {journal} {\bibinfo
  {journal} {Phys. Rev. X}\ }\textbf {\bibinfo {volume} {9}},\ \bibinfo {pages}
  {041015} (\bibinfo {year} {2019})}\BibitemShut {NoStop}%
\bibitem [{\citenamefont {Gong}\ \emph {et~al.}(2018)\citenamefont {Gong},
  \citenamefont {Ashida}, \citenamefont {Kawabata}, \citenamefont {Takasan},
  \citenamefont {Higashikawa},\ and\ \citenamefont {Ueda}}]{Gong2018}%
  \BibitemOpen
  \bibfield  {author} {\bibinfo {author} {\bibfnamefont {Z.}~\bibnamefont
  {Gong}}, \bibinfo {author} {\bibfnamefont {Y.}~\bibnamefont {Ashida}},
  \bibinfo {author} {\bibfnamefont {K.}~\bibnamefont {Kawabata}}, \bibinfo
  {author} {\bibfnamefont {K.}~\bibnamefont {Takasan}}, \bibinfo {author}
  {\bibfnamefont {S.}~\bibnamefont {Higashikawa}},\ and\ \bibinfo {author}
  {\bibfnamefont {M.}~\bibnamefont {Ueda}},\ }\bibfield  {title} {\bibinfo
  {title} {Topological phases of non-hermitian systems},\ }\href
  {https://doi.org/10.1103/physrevx.8.031079} {\bibfield  {journal} {\bibinfo
  {journal} {Phys. Rev. X}\ }\textbf {\bibinfo {volume} {8}},\ \bibinfo {pages}
  {031079} (\bibinfo {year} {2018})}\BibitemShut {NoStop}%
\bibitem [{\citenamefont {Bergholtz}\ \emph {et~al.}(2021)\citenamefont
  {Bergholtz}, \citenamefont {Budich},\ and\ \citenamefont
  {Kunst}}]{Bergholtz2021}%
  \BibitemOpen
  \bibfield  {author} {\bibinfo {author} {\bibfnamefont {E.~J.}\ \bibnamefont
  {Bergholtz}}, \bibinfo {author} {\bibfnamefont {J.~C.}\ \bibnamefont
  {Budich}},\ and\ \bibinfo {author} {\bibfnamefont {F.~K.}\ \bibnamefont
  {Kunst}},\ }\bibfield  {title} {\bibinfo {title} {Exceptional topology of
  non-hermitian systems},\ }\href
  {https://doi.org/10.1103/revmodphys.93.015005} {\bibfield  {journal}
  {\bibinfo  {journal} {Rev. Mod. Phys.}\ }\textbf {\bibinfo {volume} {93}},\
  \bibinfo {pages} {015005} (\bibinfo {year} {2021})}\BibitemShut {NoStop}%
\bibitem [{\citenamefont {Hatano}\ and\ \citenamefont
  {Nelson}(1996)}]{Hatano1996}%
  \BibitemOpen
  \bibfield  {author} {\bibinfo {author} {\bibfnamefont {N.}~\bibnamefont
  {Hatano}}\ and\ \bibinfo {author} {\bibfnamefont {D.~R.}\ \bibnamefont
  {Nelson}},\ }\bibfield  {title} {\bibinfo {title} {Localization transitions
  in non-hermitian quantum mechanics},\ }\href
  {https://doi.org/10.1103/physrevlett.77.570} {\bibfield  {journal} {\bibinfo
  {journal} {Phys. Rev. Lett.}\ }\textbf {\bibinfo {volume} {77}},\ \bibinfo
  {pages} {570} (\bibinfo {year} {1996})}\BibitemShut {NoStop}%
\bibitem [{\citenamefont {Zhang}\ \emph {et~al.}(2021)\citenamefont {Zhang},
  \citenamefont {Yang}, \citenamefont {Ge}, \citenamefont {Guan}, \citenamefont
  {Chen}, \citenamefont {Yan}, \citenamefont {Chen}, \citenamefont {Xi},
  \citenamefont {Li}, \citenamefont {Jia}, \citenamefont {Yuan}, \citenamefont
  {Sun}, \citenamefont {Chen},\ and\ \citenamefont {Zhang}}]{Zhang2021}%
  \BibitemOpen
  \bibfield  {author} {\bibinfo {author} {\bibfnamefont {L.}~\bibnamefont
  {Zhang}}, \bibinfo {author} {\bibfnamefont {Y.}~\bibnamefont {Yang}},
  \bibinfo {author} {\bibfnamefont {Y.}~\bibnamefont {Ge}}, \bibinfo {author}
  {\bibfnamefont {Y.-J.}\ \bibnamefont {Guan}}, \bibinfo {author}
  {\bibfnamefont {Q.}~\bibnamefont {Chen}}, \bibinfo {author} {\bibfnamefont
  {Q.}~\bibnamefont {Yan}}, \bibinfo {author} {\bibfnamefont {F.}~\bibnamefont
  {Chen}}, \bibinfo {author} {\bibfnamefont {R.}~\bibnamefont {Xi}}, \bibinfo
  {author} {\bibfnamefont {Y.}~\bibnamefont {Li}}, \bibinfo {author}
  {\bibfnamefont {D.}~\bibnamefont {Jia}}, \bibinfo {author} {\bibfnamefont
  {S.-Q.}\ \bibnamefont {Yuan}}, \bibinfo {author} {\bibfnamefont {H.-X.}\
  \bibnamefont {Sun}}, \bibinfo {author} {\bibfnamefont {H.}~\bibnamefont
  {Chen}},\ and\ \bibinfo {author} {\bibfnamefont {B.}~\bibnamefont {Zhang}},\
  }\bibfield  {title} {\bibinfo {title} {Acoustic non-hermitian skin effect
  from twisted winding topology},\ }\href
  {https://doi.org/10.1038/s41467-021-26619-8} {\bibfield  {journal} {\bibinfo
  {journal} {Nat. Commun.}\ }\textbf {\bibinfo {volume} {12}},\ \bibinfo
  {pages} {6297} (\bibinfo {year} {2021})}\BibitemShut {NoStop}%
\bibitem [{\citenamefont {Okuma}\ \emph {et~al.}(2020)\citenamefont {Okuma},
  \citenamefont {Kawabata}, \citenamefont {Shiozaki},\ and\ \citenamefont
  {Sato}}]{Okuma2020}%
  \BibitemOpen
  \bibfield  {author} {\bibinfo {author} {\bibfnamefont {N.}~\bibnamefont
  {Okuma}}, \bibinfo {author} {\bibfnamefont {K.}~\bibnamefont {Kawabata}},
  \bibinfo {author} {\bibfnamefont {K.}~\bibnamefont {Shiozaki}},\ and\
  \bibinfo {author} {\bibfnamefont {M.}~\bibnamefont {Sato}},\ }\bibfield
  {title} {\bibinfo {title} {Topological origin of non-hermitian skin
  effects},\ }\href {https://doi.org/10.1103/physrevlett.124.086801} {\bibfield
   {journal} {\bibinfo  {journal} {Phys. Rev. Lett.}\ }\textbf {\bibinfo
  {volume} {124}},\ \bibinfo {pages} {086801} (\bibinfo {year}
  {2020})}\BibitemShut {NoStop}%
\bibitem [{\citenamefont {Zheng}\ \emph {et~al.}(2023)\citenamefont {Zheng},
  \citenamefont {Guo}, \citenamefont {Sun}, \citenamefont {Jiang},
  \citenamefont {Li},\ and\ \citenamefont {Chen}}]{Zheng2023}%
  \BibitemOpen
  \bibfield  {author} {\bibinfo {author} {\bibfnamefont {J.}~\bibnamefont
  {Zheng}}, \bibinfo {author} {\bibfnamefont {Z.}~\bibnamefont {Guo}}, \bibinfo
  {author} {\bibfnamefont {Y.}~\bibnamefont {Sun}}, \bibinfo {author}
  {\bibfnamefont {H.}~\bibnamefont {Jiang}}, \bibinfo {author} {\bibfnamefont
  {Y.}~\bibnamefont {Li}},\ and\ \bibinfo {author} {\bibfnamefont
  {H.}~\bibnamefont {Chen}},\ }\bibfield  {title} {\bibinfo {title}
  {Topological edge modes in one-dimensional photonic artificial structures
  (invited)},\ }\href {https://doi.org/10.2528/pier22101202} {\bibfield
  {journal} {\bibinfo  {journal} {Prog. Electromagn. Res.}\ }\textbf {\bibinfo
  {volume} {177}},\ \bibinfo {pages} {1} (\bibinfo {year} {2023})}\BibitemShut
  {NoStop}%
\bibitem [{\citenamefont {Li}\ \emph {et~al.}(2023{\natexlab{b}})\citenamefont
  {Li}, \citenamefont {Zhang}, \citenamefont {Chen},\ and\ \citenamefont
  {Gao}}]{Li2023}%
  \BibitemOpen
  \bibfield  {author} {\bibinfo {author} {\bibfnamefont {Y.-Z.}\ \bibnamefont
  {Li}}, \bibinfo {author} {\bibfnamefont {Z.}~\bibnamefont {Zhang}}, \bibinfo
  {author} {\bibfnamefont {H.}~\bibnamefont {Chen}},\ and\ \bibinfo {author}
  {\bibfnamefont {F.}~\bibnamefont {Gao}},\ }\bibfield  {title} {\bibinfo
  {title} {Polarization-wavelength locked plasmonic topological states},\
  }\href {https://doi.org/10.2528/pier23081008} {\bibfield  {journal} {\bibinfo
   {journal} {Prog. Electromagn. Res.}\ }\textbf {\bibinfo {volume} {178}},\
  \bibinfo {pages} {37} (\bibinfo {year} {2023}{\natexlab{b}})}\BibitemShut
  {NoStop}%
\bibitem [{\citenamefont {Liu}\ \emph {et~al.}(2022)\citenamefont {Liu},
  \citenamefont {Gao}, \citenamefont {Wang}, \citenamefont {Xi}, \citenamefont
  {Hu}, \citenamefont {Wang}, \citenamefont {Liu}, \citenamefont {Lin},
  \citenamefont {Deng}, \citenamefont {Yang}, \citenamefont {Zhou},
  \citenamefont {Yang}, \citenamefont {Chong},\ and\ \citenamefont
  {Zhang}}]{Liu2022}%
  \BibitemOpen
  \bibfield  {author} {\bibinfo {author} {\bibfnamefont {G.-G.}\ \bibnamefont
  {Liu}}, \bibinfo {author} {\bibfnamefont {Z.}~\bibnamefont {Gao}}, \bibinfo
  {author} {\bibfnamefont {Q.}~\bibnamefont {Wang}}, \bibinfo {author}
  {\bibfnamefont {X.}~\bibnamefont {Xi}}, \bibinfo {author} {\bibfnamefont
  {Y.-H.}\ \bibnamefont {Hu}}, \bibinfo {author} {\bibfnamefont
  {M.}~\bibnamefont {Wang}}, \bibinfo {author} {\bibfnamefont {C.}~\bibnamefont
  {Liu}}, \bibinfo {author} {\bibfnamefont {X.}~\bibnamefont {Lin}}, \bibinfo
  {author} {\bibfnamefont {L.}~\bibnamefont {Deng}}, \bibinfo {author}
  {\bibfnamefont {S.~A.}\ \bibnamefont {Yang}}, \bibinfo {author}
  {\bibfnamefont {P.}~\bibnamefont {Zhou}}, \bibinfo {author} {\bibfnamefont
  {Y.}~\bibnamefont {Yang}}, \bibinfo {author} {\bibfnamefont {Y.}~\bibnamefont
  {Chong}},\ and\ \bibinfo {author} {\bibfnamefont {B.}~\bibnamefont {Zhang}},\
  }\bibfield  {title} {\bibinfo {title} {Topological chern vectors in
  three-dimensional photonic crystals},\ }\href
  {https://doi.org/10.1038/s41586-022-05077-2} {\bibfield  {journal} {\bibinfo
  {journal} {Nature}\ }\textbf {\bibinfo {volume} {609}},\ \bibinfo {pages}
  {925} (\bibinfo {year} {2022})}\BibitemShut {NoStop}%
\bibitem [{\citenamefont {Ren}\ and\ \citenamefont {Sinitsyn}(2013)}]{Ren2013}%
  \BibitemOpen
  \bibfield  {author} {\bibinfo {author} {\bibfnamefont {J.}~\bibnamefont
  {Ren}}\ and\ \bibinfo {author} {\bibfnamefont {N.~A.}\ \bibnamefont
  {Sinitsyn}},\ }\bibfield  {title} {\bibinfo {title} {Braid group and
  topological phase transitions in nonequilibrium stochastic dynamics},\ }\href
  {https://doi.org/10.1103/physreve.87.050101} {\bibfield  {journal} {\bibinfo
  {journal} {Phys. Rev. E}\ }\textbf {\bibinfo {volume} {87}},\ \bibinfo
  {pages} {050101} (\bibinfo {year} {2013})}\BibitemShut {NoStop}%
\bibitem [{\citenamefont {Yang}\ \emph {et~al.}(2019)\citenamefont {Yang},
  \citenamefont {Peng}, \citenamefont {Zhu}, \citenamefont {Buljan},
  \citenamefont {Joannopoulos}, \citenamefont {Zhen},\ and\ \citenamefont
  {Solja{\v{c}}i{\'{c}}}}]{Yang2019}%
  \BibitemOpen
  \bibfield  {author} {\bibinfo {author} {\bibfnamefont {Y.}~\bibnamefont
  {Yang}}, \bibinfo {author} {\bibfnamefont {C.}~\bibnamefont {Peng}}, \bibinfo
  {author} {\bibfnamefont {D.}~\bibnamefont {Zhu}}, \bibinfo {author}
  {\bibfnamefont {H.}~\bibnamefont {Buljan}}, \bibinfo {author} {\bibfnamefont
  {J.~D.}\ \bibnamefont {Joannopoulos}}, \bibinfo {author} {\bibfnamefont
  {B.}~\bibnamefont {Zhen}},\ and\ \bibinfo {author} {\bibfnamefont
  {M.}~\bibnamefont {Solja{\v{c}}i{\'{c}}}},\ }\bibfield  {title} {\bibinfo
  {title} {Synthesis and observation of non-abelian gauge fields in real
  space},\ }\href {https://doi.org/10.1126/science.aay3183} {\bibfield
  {journal} {\bibinfo  {journal} {Science}\ }\textbf {\bibinfo {volume}
  {365}},\ \bibinfo {pages} {1021} (\bibinfo {year} {2019})}\BibitemShut
  {NoStop}%
\bibitem [{\citenamefont {Yang}\ \emph {et~al.}(2024)\citenamefont {Yang},
  \citenamefont {Yang}, \citenamefont {Ma}, \citenamefont {Li}, \citenamefont
  {Zhang},\ and\ \citenamefont {Chan}}]{Yang2024}%
  \BibitemOpen
  \bibfield  {author} {\bibinfo {author} {\bibfnamefont {Y.}~\bibnamefont
  {Yang}}, \bibinfo {author} {\bibfnamefont {B.}~\bibnamefont {Yang}}, \bibinfo
  {author} {\bibfnamefont {G.}~\bibnamefont {Ma}}, \bibinfo {author}
  {\bibfnamefont {J.}~\bibnamefont {Li}}, \bibinfo {author} {\bibfnamefont
  {S.}~\bibnamefont {Zhang}},\ and\ \bibinfo {author} {\bibfnamefont {C.~T.}\
  \bibnamefont {Chan}},\ }\bibfield  {title} {\bibinfo {title} {Non-abelian
  physics in light and sound},\ }\href
  {https://doi.org/10.1126/science.adf9621} {\bibfield  {journal} {\bibinfo
  {journal} {Science}\ }\textbf {\bibinfo {volume} {383}},\ \bibinfo {pages}
  {eadf9621} (\bibinfo {year} {2024})}\BibitemShut {NoStop}%
\bibitem [{\citenamefont {Jiao}\ \emph {et~al.}(2021)\citenamefont {Jiao},
  \citenamefont {Longhi}, \citenamefont {Wang}, \citenamefont {Gao},
  \citenamefont {Zhou}, \citenamefont {Wang}, \citenamefont {Fu}, \citenamefont
  {Wang}, \citenamefont {Ren}, \citenamefont {Qiao},\ and\ \citenamefont
  {Jin}}]{Jiao2021}%
  \BibitemOpen
  \bibfield  {author} {\bibinfo {author} {\bibfnamefont {Z.-Q.}\ \bibnamefont
  {Jiao}}, \bibinfo {author} {\bibfnamefont {S.}~\bibnamefont {Longhi}},
  \bibinfo {author} {\bibfnamefont {X.-W.}\ \bibnamefont {Wang}}, \bibinfo
  {author} {\bibfnamefont {J.}~\bibnamefont {Gao}}, \bibinfo {author}
  {\bibfnamefont {W.-H.}\ \bibnamefont {Zhou}}, \bibinfo {author}
  {\bibfnamefont {Y.}~\bibnamefont {Wang}}, \bibinfo {author} {\bibfnamefont
  {Y.-X.}\ \bibnamefont {Fu}}, \bibinfo {author} {\bibfnamefont
  {L.}~\bibnamefont {Wang}}, \bibinfo {author} {\bibfnamefont {R.-J.}\
  \bibnamefont {Ren}}, \bibinfo {author} {\bibfnamefont {L.-F.}\ \bibnamefont
  {Qiao}},\ and\ \bibinfo {author} {\bibfnamefont {X.-M.}\ \bibnamefont
  {Jin}},\ }\bibfield  {title} {\bibinfo {title} {Experimentally detecting
  quantized zak phases without chiral symmetry in photonic lattices},\ }\href
  {https://doi.org/10.1103/physrevlett.127.147401} {\bibfield  {journal}
  {\bibinfo  {journal} {Phys. Rev. Lett.}\ }\textbf {\bibinfo {volume} {127}},\
  \bibinfo {pages} {147401} (\bibinfo {year} {2021})}\BibitemShut {NoStop}%
\bibitem [{\citenamefont {Kassel}\ and\ \citenamefont
  {Turaev}(2008)}]{Kassel2008}%
  \BibitemOpen
  \bibfield  {author} {\bibinfo {author} {\bibfnamefont {C.}~\bibnamefont
  {Kassel}}\ and\ \bibinfo {author} {\bibfnamefont {V.}~\bibnamefont
  {Turaev}},\ }\href {https://doi.org/10.1007/978-0-387-68548-9} {\emph
  {\bibinfo {title} {Braid Groups}}}\ (\bibinfo  {publisher} {Springer New
  York},\ \bibinfo {year} {2008})\BibitemShut {NoStop}%
\end{thebibliography}%

\end{document}